\documentclass[a4paper,11pt]{article}
\usepackage{jheppub}
\usepackage{braket}
\usepackage{paralist}
\usepackage{tikz}
\usepackage{float}
\usepackage{graphicx}

\usepackage[T1]{fontenc}

\title{Space-time symmetry from the world-sheet}
\author{Kiarash Naderi}
\affiliation{Institute for Theoretical Physics, ETH Zurich\\8093 Zurich, Switzerland}
\emailAdd{knaderi@phys.ethz.ch}

\abstract{The tensionless string on AdS$_3$ is a laboratory to study different aspects of the AdS/CFT correspondence. A particular question addressed in this paper is how the space-time symmetry algebra is encoded on the world-sheet. A natural prescription for calculating the space-time OPEs from the world-sheet is presented in the hybrid formalism. An argument showing that the OPEs are correctly reproduced from the world-sheet is provided, together with a few explicit calculations as consistency checks.}

\begin{document}
\maketitle
\flushbottom
%%%%%%%%%%%%%%%%%%%%%%%%%%%%%%%%%%%%%%%%%%%%%%%%%%%%
\section{Introduction and summary} \label{sec:intro}
The AdS/CFT conjecture \cite{Maldacena:1997re} is a powerful tool that relates string theory on an AdS background to a gauge theory. Although it has been checked extensively during the past decades, a first-principle derivation for most cases is still missing. One example of an AdS/CFT correspondence\- that is particularly interesting is the tensionless string on $\text{AdS}_3 \times \text{S}^3 \times \mathbb{T}^4$ with $k=1$ units of NS-NS flux. This theory is dual to the symmetric orbifold of $\mathbb{T}^4$ \cite{David:2002wn,Gaberdiel:2018rqv,Eberhardt:2018ouy}. This duality is important since there is effectively a derivation of the AdS/CFT conjecture. There have been numerous checks of this proposal, for a non-exhaustive list see \cite{Eberhardt:2018ouy,Dei:2019osr,Eberhardt:2019ywk,Eberhardt:2020akk,Knighton:2020kuh,Dei:2020zui,Eberhardt:2020bgq,Eberhardt:2021jvj,Gaberdiel:2021njm,Naderi:2022bus,Dei:2023ivl,Gaberdiel:2021kkp,Knighton:2024ybs,Fiset:2022erp,Gaberdiel:2023lco,Gaberdiel:2023dxt}. In these papers, the partition function and the world-sheet spectrum, correlation functions and deformations away from the tensionless point are matched with the symmetric orbi\-fold results. Remarkably, it has been argued that the correlation functions localize to the branched covering maps of the space-time.

The question that we would like to study in this paper is the following: how is the space-time symmetry algebra encoded on the world-sheet? A natural guess is that the world-sheet OPEs of the physical fields give rise to the space-time OPEs. This is explained in \cite{Kutasov:1999xu}. However, there are a few technicalities in applying this idea to the tensionless string mostly because of spectral flow \cite{Maldacena:2000hw,Eberhardt:2018ouy}. It is particularly interesting to understand the mechanism that localizes the world-sheet OPEs to the space-time OPEs, as one expects from the localization of the correlation functions in the tensionless limit. Moreover, the space-time theory has a huge symmetry algebra that organizes the fields, see \cite{Gaberdiel:2015mra}. It would be useful to understand how this symmetry algebra appears on the world-sheet. These are our motivations for considering the main question of this paper. In fact, as we discuss, one can overcome the technicalities and calculate the space-time OPEs directly from the world-sheet. In the remainder of this section, we give a non-technical description of the OPE proposal on the world-sheet and why we believe that it reproduces the whole space-time symmetry algebra.

Let us begin by mentioning some facts about the symmetric orbifold of $\mathbb{T}^4$, see e.g.\ \cite{Dei:2019iym} for a review. This theory contains what is called the `twisted' sectors. The single-cycle twisted sectors correspond to the conjugacy classes $[(1\cdots w)]$ with $w\geq 1$. For instance, the `untwisted' sector corresponds to $w=1$. The holomorphic symmetry algebra of $\mathbb{T}^4$ is contained in the untwisted sector of the symmetric orbifold of $\mathbb{T}^4$. This is because it is in this sector that the fields can be purely holomorphic. For this reason, in this paper we are only interested in understanding the OPEs in the untwisted sector.

Now we discuss how the OPE calculations work on the world-sheet while postponing subtleties to Section~\ref{sec:argument}. The world-sheet theory is reviewed in Section~\ref{sec:review} and Appendix~\ref{app:hybrid}. There is a physical vertex operator on the world-sheet $\mathcal{A}(x,z)$ associated to any given vertex operator $A(x)$ in the space-time. Here $z$ is the world-sheet and $x$ is the space-time coordinate which is incorporated on the world-sheet via
\begin{equation} \label{eq:intro-vertex}
	V(\ket{\mathcal{A}};x,z) = e^{x \mathcal{L}_{-1}} V(\ket{\mathcal{A}};0,z) e^{-x\mathcal{L}_{-1}} \ ,
\end{equation}
where $\mathcal{L}_{-1}$ is the space-time translation operator on the world-sheet. We will write the space-time fields in non-cursive letters while we denote the associated world-sheet physical fields by cursive letters. Let us also denote the set of physical vertex operators by $\mathcal{H}_{\text{phys.}}$. Our goal is to define a map that as an input takes two physical vertex operators and as an output gives a single physical vertex operator that corresponds to the space-time OPE
\begin{equation} \label{eq:domain-range-p}
	P: \mathcal{H}_{\text{phys.}} \times \mathcal{H}_{\text{phys.}} \rightarrow \mathcal{H}_{\text{phys.}} \ .
\end{equation}
Let us explain how this map $P$ can be defined. Assume that $\mathcal{A}(x,z)$ is physical. We begin by setting
\begin{equation} \label{eq:intro-map}
	\mathcal{A}(x) = (-1)^{f_A} \oint dz \, G^-_{-1} \mathcal{A}(x,z) \ ,
\end{equation}
where $G^-$ is the usual $b$-ghost in string theory, and $f_X=0$ ($f_X=1$) if $X$ is bosonic (fermionic). The operator in eq.~\eqref{eq:intro-map} acts on the world-sheet Fock space and maps physical states to physical states. As becomes clear in Section~\ref{sec:argument}, the additional factor $(-1)^{f_A}$ is inserted to respect certain CFT properties. Note that $\mathcal{A}(x)$ is essentially the integrated form of the vertex operator. We define $P$ by
\begin{equation} \label{eq:p-product}
	P[\mathcal{A},\mathcal{B}](x,y;w) = V(\mathcal{A}(x-y)\ket{\mathcal{B}};y,w) \ .
\end{equation}
Our claim is that $P[\mathcal{A},\mathcal{B}](x,y;w)$ gives a prescription for calculating the space-time OPE $A(x)B(y)$ from the world-sheet: if
\begin{equation}
	A(x) B(y) = \sum_{j\in \mathbb{Z}} C^{(j)}(y) (x-y)^{j} \ , \quad (|x|>|y|) \ ,
\end{equation}
then up to BRST exact terms
\begin{equation}
	P[\mathcal{A},\mathcal{B}](x,y;w) = \sum_{j\in \mathbb{Z}} \mathcal{C}^{(j)}(y,w) (x-y)^{j} \ , \quad (|x|>|y|) \ ,
\end{equation}
where $\mathcal{C}^{(j)}(y,w)$ is the physical world-sheet field that corresponds to $C(y)$.

Alternatively, one could write \eqref{eq:p-product} in terms of modes. As one expects, and we will show in Section~\ref{sec:argument}, there is a `space-time mode expansion' associated to a physical vertex operator $\mathcal{A}(x,z)$
\begin{equation} \label{eq:intro-mode}
	(-1)^{f_A} G^-_{-1} \mathcal{A}(x,z) = \sum_{r\in\mathbb{Z}-h_A} \mathcal{A}_r(z) \, x^{-r-h_A} \ ,
\end{equation}
where $h_A$ is the space-time conformal dimension of $\mathcal{A}(x,z)$. Similar to eq.~\eqref{eq:intro-map}, if we define
\begin{equation} \label{eq:intro-def-mode}
	\mathcal{A}_r = \oint dz \, \mathcal{A}_r(z) \ ,
\end{equation}
this operator maps physical states to physical states. Using this, \eqref{eq:p-product} can be written as
\begin{equation} \label{eq:product-modes}
	P[\mathcal{A},\mathcal{B}](x,y;w) = \sum_{r\in\mathbb{Z}-h_A} V(\mathcal{A}_r \ket{\mathcal{B}};y,w) \, (x-y)^{-r-h_A} \ .
\end{equation}
The map $P$ that we just discussed can be evaluated for two physical fields and can be compared with the space-time OPEs. We perform such consistency checks in Section~\ref{sec:checks} and in Appendix~\ref{app:example} using the formulation of \cite{Dei:2023ivl}. In particular, we expli\-citly calculate the OPE of two supercurrents and the OPE of the stress-tensor with itself. However, in the tensionless string, one can actually provide an argument that this structure correctly reproduces the space-time OPEs, as we summarize in the remainder of this section. Eq.~\eqref{eq:product-modes} is how one calculates OPEs in a CFT, see e.g.\ \cite{Gaberdiel:1998fs,Kac:1996wd}. Therefore, if we show that $\mathcal{A}_r \ket{\mathcal{B}}$ corresponds to the space-time state $A_r\ket{B}$, this implies that the OPE in eq.~\eqref{eq:product-modes} coincides with the space-time OPE. We will argue for this recursively using specific information about the spectrum of the space-time theory.

In the untwisted sector of the space-time theory, any state is generated by the modes of the following fields
\begin{itemize}
	\item the symmetrized free bosons $\partial X^j$, $\partial \bar{X}^j$ with $j\in\{1,2\}$,
	\item the symmetrized free fermions $\psi^{\alpha,j}$ with $\alpha\in\{+,-\}$ and $j\in\{1,2\}$.
\end{itemize}
They satisfy the algebra
\begin{equation} \label{eq:intro-st}
	[\partial X^j_n,\partial \bar{X}^k_m] = n \delta^{jk} \delta_{n+m,0} \ , \quad \{\psi^{\alpha,j}_r,\psi^{\beta,k}_s\} = \epsilon^{\alpha\beta} \epsilon^{kj} \delta_{r+s,0}\ .
\end{equation}
We collectively denote these fields as $F$. Therefore, a given state $\ket{A}$ can be written as a linear sum of the states of the following form with $s_j>0$\footnote{In this paper, we only consider vertex operators with zero momenta and winding numbers in the untwisted sector, see e.g.\ \cite{Naderi:2022bus} for a comment on this.\label{footnote:momenta-winding}}
\begin{equation} \label{eq:intro-st-spectrum}
	\ket{A}=F^{(1)}_{-s_1} \cdots F^{(n)}_{-s_n} \ket{0} \ .
\end{equation}
$\ket{0}$ is the space-time vacuum. On the world-sheet, we denote the physical fields that correspond\- to $F$ as $\mathcal{F}$. As we show in Section~\ref{sec:checks}, the space-time modes of these fields, see eq.~\eqref{eq:intro-def-mode}, satisfy the space-time algebra given in eq.~\eqref{eq:intro-st}. Also note that they map physical states to physical states. For a given state $\ket{A}$ of the form in eq.~\eqref{eq:intro-st-spectrum}, we define the corresponding physical state $\ket{\mathcal{A}}$ as
\begin{equation} \label{eq:intro-spectrum}
	\ket{\mathcal{A}} = \mathcal{F}^{(1)}_{-s_1} \cdots \mathcal{F}^{(n)}_{-s_n} \ket{\Omega} \ ,
\end{equation}
where $\ket{\Omega}$ is the physical state that corresponds to $\ket{0}$. In order to show that $A_r \ket{B}$ corresponds\- to $\mathcal{A}_r \ket{\mathcal{B}}$, we write the states $\ket{B}$ and $\ket{\mathcal{B}}$ as in eqs.~\eqref{eq:intro-st-spectrum} and \eqref{eq:intro-spectrum} respectively. This implies that it is enough to show that $A_r \ket{0}$ corresponds to $\mathcal{A}_r \ket{\Omega}$, and also that $[A_r,F_s]$ corresponds to $[\mathcal{A}_r,\mathcal{F}_s]$. As expected and shown in Section~\ref{sec:argument}, we have
\begin{equation} \label{eq:intro-loc}
	\mathcal{A}_r \ket{\Omega} = 0 \ , \quad (r>-h_A) \ , \quad \mathcal{A}_{-h_A} \ket{\Omega} = \ket{\mathcal{A}} \ .
\end{equation}
We show in Appendix~\ref{app:ope}, essentially from the definition in eq.~\eqref{eq:intro-vertex}, that eq.~\eqref{eq:intro-loc} implies that $A_r \ket{0}$ corresponds to $\mathcal{A}_r \ket{\Omega}$ for any $r\in\mathbb{Z}-h_A$. In order to argue for the agreement of the algebras, we show in Section~\ref{sec:argument} that $[\mathcal{A}_n,\mathcal{B}_m]$ can be written as the usual contour integral of the corresponding space-time OPE in eq.~\eqref{eq:product-modes}, see eq.~\eqref{eq:anti-comm-s-t}. Therefore, for calculating $[\mathcal{F}_s,\mathcal{A}_r]$, we write the singular part of the OPE of $\mathcal{F}(x,z)$ with $\mathcal{A}(y,w)$ written as in eq.~\eqref{eq:intro-spectrum}. As we mentioned, the space-time modes of the free bosons and free fermions satisfy the algebra in eq.~\eqref{eq:intro-st} on the world-sheet. Using this and eqs.~\eqref{eq:intro-spectrum} and \eqref{eq:intro-loc}, we see that the singular part of the OPE in eq.~\eqref{eq:product-modes} agrees with the space-time result. This implies that $[\mathcal{A}_r,\mathcal{F}_s]$ matches with $[A_r,F_s]$, and we conclude that the OPE prescription in eq.~\eqref{eq:product-modes} agrees with the space-time OPE prescription.

In Section~\ref{sec:w-algebra}, we discuss how the space-time $\mathcal{W}_{\infty}$ algebra arises from the world-sheet point of view. Section~\ref{sec:conclusion} includes our conclusions and possible future directions. There are a few appendices to make the paper self-contained.
%%%%%%%%%%%%%%%%%%%%%%%%%%%%%%%%%%%%%%%%%%%%%%%%%%%%%%%%%%%%%%%%%%%%%%
\section{A review of the tensionless string} \label{sec:review}
In this section, we briefly discuss necessary tools that are needed to follow the paper in detail: we review the tensionless string on a specific $\text{AdS}_3$ background. A familiar reader may briefly see Section~\ref{sec:psu} and skip the rest of this section, and only look at the equations that are referred to later on.

The string theory that we are interested in is the superstring with one unit of NS-NS flux on the background $\text{AdS}_3\times \text{S}^3\times \mathbb{T}^4$. The size of $\text{AdS}_3$ and $\text{S}^3$ equals $k=1$ in string units \cite{Eberhardt:2018ouy}. This string theory is well-defined in the so-called 'hybrid formalism' of Berkovits, Vafa and Witten (BVW) \cite{Berkovits:1999im}. The hybrid formalism and its relevant ingredients for our purposes are briefly reviewed in Appendix~\ref{app:hybrid}. In the following, we sketch the fields that appear in this formalism and the physical state conditions.

The hybrid construction makes the space-time supersymmetry of $\text{AdS}_3 \times \text{S}^3$ manifest on the world-sheet while $\mathbb{T}^4$ is still described in the RNS formalism \cite{Green:1987sp}. More specifically, in the superstring theory, we still need the $bc$- and $\beta\gamma$-ghosts. After a set of field redefinitions \cite{Berkovits:1999im}, the superstring theory is described by the following (anti-)commuting pieces\footnote{In the following, we will mainly focus on the holomorphic part of the world-sheet theory.}
\begin{itemize}
	\item a WZW model on $\mathfrak{psu}(1,1|2)_1$ with $c=-2$,
	\item a boson $\sigma$ which is the bosonization of the usual $bc$-system with $c=-26$,
	\item a boson $\rho$ which is a combination of the $\beta\gamma$-ghost and other fields with $c=+28$,
	\item a topologically twisted theory on $\mathbb{T}^4$, see Appendix~\ref{app:t4} for our conventions, with $c=0$.
\end{itemize}
We note that the central charges add up to $c=0$, which is required for a critical world-sheet theory \cite{Green:1987sp}. It is non-trivial to check that these theories combine to realize a topologically twisted $\mathcal{N}=4$ algebra on the world-sheet with $c=6$ \cite{Berkovits:1993xq,Berkovits:1994vy,Berkovits:1999im}, see Appendix~\ref{app:hybrid}. We review our conventions for the OPEs of an $\mathcal{N}=4$ algebra and its topologically twisted version in Appendix~\ref{app:sca}. Let us specify the fields of a topologically twisted $\mathcal{N}=4$ algebra
\begin{itemize}
	\item the stress-tensor $T$ (primary),
	\item the R-symmetry currents $J$ (non-primary\footnote{Of course if $c\neq 0$.}), $J^{++}$ and $J^{--}$ (primary) which together realize $\mathfrak{su}(2)_1$,
	\item the supercurrents $G^+$, $G^-$, $\tilde{G}^+$ and $\tilde{G}^-$ (primary).
\end{itemize}
We note that the number of '$\pm$' indicates twice the charge under $J$, e.g.\ $J^{++}$ has charge $1$. Also, $T$, $J^{--}$, $G^-$ and $\tilde{G}^-$ have weight $2$; $G^+$, $\tilde{G}^+$ and $J$ have weight $1$; and $J^{++}$ has weight $0$, see Appendix~\ref{app:sca}.

As this is a world-sheet theory, one needs to specify the physical state conditions. It turns out that the physical states are described by a (double-)cohomology as follows
\begin{equation} \label{eq:physical}
	G^+_0 \psi = \tilde{G}^+_0 \psi = L_0 \psi = (J_0-\tfrac{1}{2})\psi = 0 \ , \quad \psi \sim \psi + G^+_0 \tilde{G}^+_0 \phi \ ,
\end{equation}
where $L_0$ is the zero mode of the total world-sheet stress-tensor, see eq.~\eqref{eq:hybrid-stress-tensor}. $G^+_0$ plays the role of the BRST operator, and the $\tilde{G}^+_0 \psi=0$ condition ensures the restriction to the small Hilbert space.

In the tensionless limit where $k=1$, there is a free-field realization for the WZW model on $\mathfrak{psu}(1,1|2)_1$ \cite{Eberhardt:2018ouy,Dei:2020zui} in terms of symplectic bosons and free fermions \cite{Lesage:2002ch,Gaiotto:2017euk}. For our purpose, it is more convenient to work with an alternative realization for $\mathfrak{psu}(1,1|2)_1$ proposed in \cite{Beem:2023dub} and also in \cite{Dei:2023ivl}.
%%%%%%%%%%%%%%%%%%%%%%%%%%%%%%%%%%%%%%%%%%%%%%%%%%%%%%%%%%%%%%%%%%%%%
\subsection{An alternative realization for \texorpdfstring{$\mathfrak{psu}(1,1|2)_1$}{psu(1,1|2)1}} \label{sec:psu}
In the formulation of \cite{Beem:2023dub,Dei:2023ivl}, the $\mathfrak{psu}(1,1|2)_1$ theory is described in terms of a $\beta\gamma$-system and $2$ $bc$-systems which we denote as follows\footnote{The $\beta\gamma$-system here should not be confused with the superdiffeomorphism ghosts. In fact, we never use the latter ghosts directly and we use instead the variables of the hybrid formalism listed in this section.}
\begin{equation} \label{eq:realization}
	(\beta,\gamma) \ , \quad (p_j,\theta^j) \ , \quad j\in\{1,2\} \ .
\end{equation}
They have the weights
\begin{equation} \label{eq:weights}
	\Delta(\beta)=\Delta(p_j)=1 \ , \quad \Delta(\gamma)=\Delta(\theta^j)=0 \ ,
\end{equation}
and satisfy the OPEs
\begin{equation}
	\beta(z) \gamma(w) \sim \frac{-1}{(z-w)} \ , \quad p_j(z) \theta^k(w) = \frac{\delta_j^k}{(z-w)} \ .
\end{equation}
One can write down the $\mathfrak{psu}(1,1|2)_1$ currents in terms of these variables as\footnote{This implies that only the $R$-sector of the fields in \eqref{eq:realization} should be considered because the $\mathfrak{psu}(1,1|2)_1$ currents are integer-moded.}
\begin{subequations} \label{eq:psu-currents}
	\begin{equation}
		J^+ = \beta \ , \quad J^3 = (\beta\gamma)+\frac{1}{2}(p_j\theta^j) \ , \quad J^- = (\beta\gamma)\gamma+(p_j\theta^j)\gamma \ ,
	\end{equation}
	\begin{equation}
		K^+ = (p_2 \theta^1) \ , \quad K^3 = \frac{\epsilon^{ij}}{2}(p_j\theta^j) \ , \quad K^- = (p_1\theta^2) \ ,
	\end{equation}
	\begin{equation}
		S^{+++}=p_2 \ , \quad S^{+-+}=p_1 \ , \quad S^{-++}=-(\gamma p_2) \ , \quad S^{---}=[(\beta\gamma)+(p_j\theta^j)]\theta^2 \ ,
	\end{equation}
	\begin{equation}
		S^{++-}=(\beta\theta^1) \ , \quad S^{+--}=-(\beta\theta^2) \ , \quad S^{--+}=-(\gamma p_1) \ , \quad S^{-+-}=-[(\beta\gamma)+(p_j\theta^j)]\theta^1 \ .
	\end{equation}
\end{subequations}
In particular, the space-time M\"{o}bius transformations are realized on the world-sheet by the zero modes of the $\mathfrak{sl}(2,\mathbb{R})_1$ currents $J^a$, and the global part of the R-symmetry currents is realized by the zero modes of $K^a$ which form an $\mathfrak{su}(2)_1$. The zero-mode representation of the fields in \eqref{eq:realization} is
\begin{equation}
	\beta_0 \ket{m} = m\ket{m+1} \ , \quad \gamma_0 \ket{m}=\ket{m-1} \ , \quad (p_j)_0\ket{m}=0 \ , \quad j\in\{1,2\} \ .
\end{equation}
There is an automorphism for the realization in eqs.~\eqref{eq:psu-currents} called spectral flow \cite{Maldacena:2000hw,Eberhardt:2018ouy,Dei:2023ivl}. The automorphism $\sigma^w$ is defined for $w\in \mathbb{Z}$ by\footnote{In the following, we only consider the case $w\geq 1$. The case of negative $w$'s is interpreted as the conjugation of the spectrally flowed states, also see \cite{Eberhardt:2018ouy}.}
\begin{subequations} \label{eq:spectral-flow}
\begin{equation}
	\sigma^w(\beta_n)=\beta_{n-w} \ , \quad \sigma^w(\gamma_n) = \gamma_{n+w} \ ,
\end{equation}
\begin{equation}
	\sigma^w((p_1)_n)=(p_1)_{n-w} \ , \quad \sigma^w(\theta^1_n) = \theta^1_{n+w} \ ,
\end{equation}
\begin{equation}
	\sigma^w((p_2)_n)=(p_2)_{n} \ , \quad \sigma^w(\theta^2_n) = \theta^2_{n} \ .
\end{equation}
\end{subequations}
This defines the $w$-spectrally flowed representations via
\begin{equation} \label{eq:def-spectral-flow-action}
	F_n [\ket{m}]^{\sigma^w} = [\sigma^w(F_n)\ket{m}]^{\sigma^w} \ .
\end{equation}
In other words, the vector space of the representations is the same while the action of the generators is modified. For later applications, it is useful to bosonize \cite{Friedan:1985ge,Friedan:1986rx,Naderi:2022bus,Dei:2023ivl}
\begin{subequations} \label{eq:bosonization}
\begin{equation}
	\theta^1 = e^{if_1} \ , \quad p_1 = e^{-if_1} \ , \quad \theta^2=e^{-if_2} \ , \quad p_2 = e^{if_2} \ .
\end{equation}
\end{subequations}
They satisfy
\begin{equation}
	A(z)A(w)\sim -\ln{(z-w)} \ , \quad A\in\{f_1,f_2\} \ ,
\end{equation}
while they have trivial OPEs between each other. The background charges of $f_1$ and $f_2$ are chosen such that the weights are as in eq.~\eqref{eq:weights}, see eq.~\eqref{eq:psu-stress-tensor}. We have written the generators of the world-sheet topologically twisted $\mathcal{N}=4$ algebra in terms of the realization in \eqref{eq:realization}, see Appendix~\ref{app:hybrid}.

After this detour, the question is what should be done to `solve' the string theory that we have described so far. The first task is to find the physical states that satisfy eq.~\eqref{eq:physical}. The next task is to consider the correlation functions of these vertex operators on different world-sheet topologies. The former is discussed in \cite{Eberhardt:2018ouy,Giveon:1998ns,Eberhardt:2019qcl,Naderi:2022bus,Eberhardt:2020bgq,Eberhardt:2021jvj} while the latter is done in \cite{Eberhardt:2019ywk,Dei:2020zui,Dei:2023ivl,Eberhardt:2020akk,Knighton:2020kuh}. We discuss these two tasks in turn below.
%%%%%%%%%%%%%%%%%%%%%%%%%%%%%%%%%%%%
\subsection{The space-time spectrum} \label{sec:spectrum}
The question that we intend to answer in this subsection is the following: how is the space-time spectrum generated from the world-sheet? Let us begin by reviewing the spectrum of the space-time theory. The (single-cycle) states of the space-time theory are described by the twisted sectors that correspond to conjugacy classes $[(1 \cdots w)]$ for $w\geq 1$. The space-time spectrum is then generated by applications of the fractionally-moded free fermions and free bosons of $\mathbb{T}^4$ on the $w$-twisted ground states, see e.g.\ \cite{Dei:2019iym} for a review.\footnote{For the untwisted sector, one considers the symmetrized operators, see e.g.\ eq.~(2.7) in \cite{Fiset:2022erp}.} In particular, for $w$-odd, we denote the vertex operator of the $w$-twisted ground state by $\sigma_w$, while for $w$-even, we denote them by $\sigma_w^{\pm}$.\footnote{For $w$-even, there are $2$ more fields as there are $4$ real fermions, see e.g.\ \cite{Gaberdiel:2015uca}. Here we focus on the two fields that make a doublet with respect to the $\mathfrak{su}(2)_1$ of $\mathfrak{psu}(1,1|2)_1$, see \cite{Gaberdiel:2021njm}.} From the world-sheet perspective, a combination of these fields that respect orbifold invariance should be physical. In \cite{Dei:2020zui}, the physical fields that correspond to the $w$-twisted ground states on the world-sheet were found. In fact, in \cite{Eberhardt:2018ouy}, it was shown that the $w$-twisted sector corresponds to the $w$-spectrally flowed representation on the world-sheet, see eq.~\eqref{eq:def-spectral-flow-action}. We will denote the world-sheet vertex operators that correspond to the $w$-twisted ground states by $\Omega_w$ for $w$-odd, and by $\Omega_w^{\pm}$ for $w$-even. We have \cite{Dei:2023ivl}\footnote{Note that we have put a minus sign in front of the vertex operators compared to \cite{Naderi:2022bus,Dei:2023ivl}. This is because it seems more natural for us to set the normalization to $1$ in the picture $P=0$ rather than $P=-2$.\label{footnote:normalization}}
\begin{subequations} \label{eq:w-twisted-ground-states}
	\begin{equation} \label{eq:w-odd}
		\Omega_w(x,z) = -\exp{\Big[\frac{w+1}{2}(if_1-if_2)\Big]} \Big( \frac{\partial^w \gamma}{w!} \Big)^{-m_w} \delta_w(\gamma(z)-x) e^{2\rho+i\sigma+iH} \ ,
	\end{equation}
	\begin{equation} \label{eq:w-even}
		\Omega_w^{\pm}(x,z) = -\exp{\Big[\frac{\pm (if_1+if_2)}{2} \Big]} \exp{\Big[\frac{w+1}{2}(if_1-if_2)\Big]} \Big( \frac{\partial^w \gamma}{w!} \Big)^{-m_w^{\pm}} \delta_w(\gamma(z)-x) e^{2\rho+i\sigma+iH} \ ,
	\end{equation}
	\begin{equation} \label{eq:mw}
		m_w = -\frac{(w-1)^2}{4w}\ , \quad m_w^{\pm}=-\frac{w-2}{4} \ .
	\end{equation}
\end{subequations}
A few comments are in order: 1) the variable $x$ is the space-time coordinate which is incorporated in the world-sheet vertex operators via
\begin{equation} \label{eq:vertex-operator-x-z}
	V(\phi;x,z)=e^{x J^+_0} e^{z L_{-1}} V(\phi;0,0) e^{-z L_{-1}} e^{-x J^+_0} \ ,
\end{equation}
for a state $\phi$ on the world-sheet. Note that $J^+_0$ is the space-time and $L_{-1}$ is the world-sheet translation operator, and they commute; 2) the delta-functions are defined as
\begin{equation} \label{eq:w>1-delta}
	\delta_w(\gamma(z)-x) = \delta(\gamma(z)-x) \prod_{j=1}^{w-1} \delta(\partial^j \gamma(z)) \ .
\end{equation}
As we will see in the subsequent sections, this in particular means that we describe $(p_j,\theta^j)$ in the operator-formalism while $(\beta,\gamma)$ is described in the path-integral formalism \cite{Dei:2023ivl}. This reproduces the expected OPEs from the path-integral perspective, see \cite{Verlinde:1987sd,Witten:2012bh}. Moreover, as we discuss in Section~\ref{sec:argument} and in Appendix~\ref{app:vertex}, (at least) in the case of $w=1$, one can make sense of the delta-functions in the language of vertex algebras \cite{Kac:1996wd}. This is the case that we consider in this paper, since as we mentioned in the Introduction, we are interested in the untwisted sector of the space-time theory; 3) the fields in eqs.~\eqref{eq:w-twisted-ground-states} are physical, i.e.\ they satisfy the conditions in eq.~\eqref{eq:physical}; 4) as we are in a superstring theory with superdiffeomorphism ghosts, the physical states have a picture-number and can be equally expressed in different pictures, see \cite{Friedan:1985ey,Friedan:1985ge,Friedan:1986rx,Blumenhagen:2013fgp,Lust:1989tj}. We have briefly reviewed this in Appendix~\ref{app:hybrid}. The picture number of a vertex operator is its eigenvalue under the current $P_{\rho}=\partial \rho$. This for instance means that the fields in eqs.~\eqref{eq:w-twisted-ground-states} have picture number $P=-2$. One can equivalently write these fields in other pictures, as we will briefly mention below; 5) finally, $H^j$ and $H$ are defined in Appendix~\ref{app:t4}, see eq.~\eqref{eq:def-h}.

Up to now we have discussed the world-sheet states that correspond to the space-time $w$-twisted ground states. As we mentioned, the space-time spectrum is generated by acting with fractionally-moded free bosons and free fermions on these states. Associated to each mode of the free bosons and free fermions, there is a Del~Giudice, Di~Vecchia and Fubini (DDF) operator \cite{DelGiudice:1971yjh} on the world-sheet. DDF operators are operators that commute with the operators that define the physical state conditions in eq.~\eqref{eq:physical}, and so they map physical states to physical states. Therefore, the space-time spectrum is generated by iterative applications of the DDF operators of the free bosons and free fermions on the $w$-twisted ground states \cite{Naderi:2022bus}, also see footnote~\ref{footnote:momenta-winding}. In particular, this means that the normalization of a given state on the space-time is fixed on the world-sheet once the normalization of the $w$-twisted ground states are fixed in a given picture which we choose to be $P=0$, see eq.~\eqref{eq:ground-state-pictures}. We will discuss the spectrum for the case of $w=1$ in more detail in Section~\ref{sec:argument}. For later applications, we write the DDF operators associated with the free bosons and the free fermions of $\mathbb{T}^4$ \cite{Naderi:2022bus}
\begin{subequations} \label{eq:ddf-ops}
	\begin{equation} \label{eq:ddf-bosons}
		\partial \bar{\mathcal{X}}^j_n =\oint dz \, \partial \bar{X}^j \gamma^n \ , \quad \partial \mathcal{X}^j_n = \oint dz \, \Big[\partial X^j \gamma^n -n \gamma^{n-1} e^{\rho+iH^j} e^{if_1 - if_2} \Big] \ ,
	\end{equation}
	\begin{equation} \label{eq:ddf-fermions}
		\Psi^{+,j}_r=\oint dz \, e^{if_1} \gamma^{r-\frac{1}{2}} e^{\rho+iH^j} \ , \quad \Psi^{-,j}_r=-\oint dz \, e^{-if_2} \gamma^{r-\frac{1}{2}} e^{\rho+iH^j} \ ,
	\end{equation}
\end{subequations}
where $j\in \{1,2\}$. Note that $n\in\mathbb{Z}$ and $r\in\frac{1}{2}+\mathbb{Z}$ as the space-time fermions are in the NS-sector. Throughout the text, we absorb the $1/(2\pi i)$ factor inside the definition of contour integrals. In \cite{Naderi:2022bus}, it was shown that
\begin{equation} \label{eq:ddf-algebra}
	[\partial \mathcal{X}^j_n,\partial \bar{\mathcal{X}}^k_m]=n\delta^{jk} \delta_{n+m,0} \mathcal{I}\ , \quad \{\Psi^{\alpha,j}_r,\Psi^{\beta,l}_s\} = \epsilon^{\alpha\beta} \epsilon^{lj} \delta_{r+s,0} \mathcal{I} \ ,
\end{equation}
where $\mathcal{I}$ is the `identity' operator defined as \cite{Giveon:1998ns,Eberhardt:2019qcl}
\begin{equation} \label{eq:ddf-identity-operator}
	\mathcal{I}=\oint dz \, \gamma^{-1}\partial\gamma \ .
\end{equation}
This is in agreement with the space-time algebra, see eq.~\eqref{eq:bos-ferm-t4-ope}. Note that $\mathcal{I}$ commutes with the mentioned DDF operators. We mention in passing that it will prove useful to consider the $w=1$-twisted ground states in different pictures. For this case, we have (for $n\geq 0$)\footnote{For brevity, since $w=1$ is assumed, here and below we drop the $w$-subscript. Instead, we specify the picture number by a subscript $P$. Also see footnote~\ref{footnote:normalization}.}
\begin{equation} \label{eq:ground-state-pictures}
	\Omega_{P=-2n}(x,z) = (-1)^{n} e^{inf_1-inf_2} (\partial \gamma)^{1-n} \delta(\gamma(z)-x) e^{2n\rho+i\sigma+inH} \ .
\end{equation}
%%%%%%%%%%%%%%%%%%%%%%%%%%%%%%%%%%%%%%%%%%%%%%%%%%
\subsection{Correlation functions} \label{sec:correlation}
The space-time theory is the symmetric orbifold of $\mathbb{T}^4$. The correlation functions of the symmetric orbifold are discussed e.g.\ in \cite{Lunin:2000yv,Lunin:2001pw,Pakman:2009zz,Roumpedakis:2018tdb,Dei:2019iym}. For concreteness, let us focus on the case where the space-time is $\mathbb{C}\mathbb{P}^1$. In this case, the correlation functions of $w$-twisted ground states are described as a sum over branched covering maps from the world-sheet to the space-time. In \cite{Eberhardt:2019ywk,Dei:2020zui,Dei:2023ivl}, it was shown that one can calculate the world-sheet tree-level correlation functions of $w$-twisted ground states and recover the space-time result. As we will see below, there is a very natural connection between our prescription for calculating the space-time OPEs and the world-sheet correlation functions. For this reason, we briefly review the prescription for calculating tree-level correlation functions \cite{Berkovits:1994vy}.

Consider a set of physical vertex operators $V_j$ with $j\in\{1,\cdots,n\}$. We fix $3$ points $z_i$ and also $x_i$. The tree-level amplitude is then \cite{Berkovits:1994vy}
\begin{equation} \label{eq:tree-level}
	\mathcal{I} = \int d^2 z_4 \cdots d^2 z_n \bigl|\bigl<\tilde{V}_1(x_1,z_1) V_2(x_2,z_2) V_3(x_3,z_3) \prod_{j=4}^n G^-_{-1} V_j(x_j,z_j)\bigr>\bigr|^2 \ ,
\end{equation}
where $G^-$ is the usual $b$-ghost, see Appendix~\ref{app:hybrid}. In other words, $3$ vertex operators are unintegrated while the rest are integrated. The vertex operator $\tilde{V}_1$ is defined as a pre-image of $V_1$ under $\tilde{G}^+_0$ \cite{Berkovits:1994vy}
\begin{equation}
	V_1 = \tilde{G}^+_0 \tilde{V}_1 \ .
\end{equation}
Note that the operation $G^+_0 \tilde{V}_1$ is (up to BRST exact states) the picture-raising operator $P_+$ defined in Appendix~\ref{app:hybrid} \cite{Friedan:1985ey,Friedan:1986rx,Blumenhagen:2013fgp,Lust:1989tj}. For this reason, eq.~\eqref{eq:tree-level} agrees with the prescription used in \cite{Dei:2023ivl} if the sum of picture numbers is assumed to be $\sum_{j=1}^n P_j = -4$, see \cite{Berkovits:1994vy}. In fact, $\tilde{V}_1$ is needed to balance the background charges of $\rho$, $\sigma$ and $H$, see Appendix~\ref{app:hybrid}.

This prescription in particular implies that the integrated vertex operator takes the following form
\begin{equation} \label{eq:integrated-vertex-operator}
	\overline{V}(x,\bar{x})=\int d^2 z \Big[G^-_{-1} V(x,z;\bar{x},\bar{z}) \Big] \ .
\end{equation}
We will show below that once we consider the world-sheet OPE of an `integrated' vertex operator with a physical field, this reproduces the corresponding space-time OPE.
%%%%%%%%%%%%%%%%%%%%%%%%%%%%%%%%%%%%%%%%%%%%%%%%%%%%%%%%%%%%%%%%%%%%%%%
\section{The space-time OPEs from the world-sheet} \label{sec:argument}
In this section, we discuss in more detail the prescription for calculating the space-time OPEs from the world-sheet. We argue that this structure reproduces the expected space-time results in the tensionless string. Our aim is to keep this section self-contained while discussing certain subtleties that we did not cover in the Introduction.

As we mentioned, the idea is to define a map $P$ that takes two physical vertex operators and gives the space-time OPE associated to them. We begin by discussing how this map can be defined. Given a vertex operator $A(x)$ in the untwisted sector (denoted by non-cursive letters), there is a corresponding physical vertex operator $\mathcal{A}(x,z)$ on the world-sheet (denoted by cursive letters), i.e.\ it satisfies eq.~\eqref{eq:physical}. This statement follows from the discussion in Section~\ref{sec:spectrum}. We denote the set of physical vertex operators by $\mathcal{H}_{\text{phys.}}$. It is then straightforward to see that the following operator\footnote{Throughout the text, we use either $\mathcal{A}(x,z)$ or $V(\ket{\mathcal{A}};x,z)$ equivalently for a physical vertex operator. The same holds between $\mathcal{A}(x)$ and $W(\ket{\mathcal{A}},x)$.}
\begin{equation} \label{eq:arg-ax}
	\mathcal{A}(x) = W(\ket{\mathcal{A}},x) = (-1)^{f_A} \oint dz \, V(G^-_{-1} \ket{\mathcal{A}};x,z) \ ,
\end{equation}
is a DDF operator if $\ket{\mathcal{A}}$ is physical. Therefore, $\mathcal{A}(x)$ maps physical states to physical states. Note that $G^-$ is the usual $b$-ghost in string theory, see Appendix~\ref{app:hybrid}. The factor $(-1)^{f_A}$ is inserted to respect certain properties, e.g.\ see eq.~\eqref{eq:locality-assumption}, where $f_X$ denotes the world-sheet parity of $V(G^-_{-1} \ket{\mathcal{X}};x,z)$, see Appendix~\ref{app:ope}. The operator $\mathcal{A}(x)$ is the zero mode of the vertex operator $(-1)^{f_A}V(G^-_{-1} \ket{\mathcal{A}};x,z)$. We define $P: \mathcal{H}_{\text{phys.}} \times \mathcal{H}_{\text{phys.}} \rightarrow \mathcal{H}_{\text{phys.}}$ by
\begin{equation} \label{eq:p-product-arg}
	P[\mathcal{A},\mathcal{B}](x,y;w) = V(\mathcal{A}(x-y)\ket{\mathcal{B}};y,w) \ .
\end{equation}
In the space-time theory we have
\begin{equation}
	A(x) B(y) = \sum_{j\in\mathbb{Z}} C^{(j)}(y) (x-y)^{j} \ , \quad (|x|>|y|) \ ,
\end{equation}
for two fields $A(x)$ and $B(y)$. On the world-sheet, associated to each field $C^{(j)}(y)$, there is a physical vertex operator $\mathcal{C}^{(j)}(y,w)$. Then the claim is that
\begin{equation}
	P[\mathcal{A},\mathcal{B}](x,y;w) = \sum_{j\in\mathbb{Z}} \mathcal{C}^{(j)}(y,w) (x-y)^{j} \ , \quad (|x|>|y|) \ ,
\end{equation}
up to BRST exact terms. In Appendix~\ref{app:ope}, we study some properties of the prescription in eq.~\eqref{eq:p-product-arg}, for instance, we show that
\begin{equation} \label{eq:explicit-p-formula}
	P[\mathcal{A},\mathcal{B}](x,y;w) = (-1)^{f_A} \oint_{C_w} dz \, [G^-_{-1} \mathcal{A}(x,z)] \mathcal{B}(y,w) \ ,
\end{equation}
where $C_w$ is a small counter-clockwise circle around $w$. The prescription that we just discussed can in principle be calculated for two phy\-sical fields and can be compared to the corresponding space-time OPE. However, in the tensionless\- string, one can provide an argument that this structure correctly reproduces the space-time OPEs. Before delving into more detail, let us broadly explain how the argument\- works. The general idea is to express eq.~\eqref{eq:p-product-arg} in terms of `space-time modes', which we will define below. This will eventually allow us to write the map $P$ in terms of (anti-)commutators of these modes. Once we show that these modes satisfy the expected space-time algebra, we deduce that our prescription reproduces the space-time OPEs. We will now focus on explaining our argument in detail.

We begin by discussing the space-time mode expansion of a physical vertex operator. For a physical vertex operator $\mathcal{A}(x,z)$, one expects to be able to define the corresponding space-time mode expansion
\begin{equation} \label{eq:mode-expansion}
	(-1)^{f_A} V(G^-_{-1}\ket{\mathcal{A}};x,z) = \sum_{r\in \mathbb{Z}-h_A} \mathcal{A}_r(z) \, x^{-r-h_A} \ .
\end{equation}
Here we write $h_F$ as the space-time weight of a field $\mathcal{F}$, also see eq.~\eqref{eq:arg-ax}. Moreover, one expects that the zero mode of $\mathcal{A}_r(z)$ is a combination of the modes of free bosons and free fermions. However, in order to concretely realize these expectations on the world-sheet, we will take a short detour.

As we discussed in Section~\ref{sec:spectrum}, any physical field $\mathcal{A}(x,z)$ which corresponds to a state in the untwisted sector (with zero momenta and winding numbers) can be written as a linear sum of the vertex operators of the form \cite{Eberhardt:2018ouy,Naderi:2022bus}
\begin{equation} \label{eq:generic-state-arg}
	V(\ket{\mathcal{A}};x,z) = V(\mathcal{F}^{(1)}_{-s_1} \cdots \mathcal{F}^{(n)}_{-s_n} \ket{\Omega_{P=0}};x,z) \ ,
\end{equation}
for some $n\geq 0$ and $s_j\in\mathbb{Z}_{\geq 1}$ if bosonic and $s_j\in \frac{1}{2}+\mathbb{Z}_{\geq 0}$ if fermionic. $\mathcal{F}^{(j)}_{-s_j}$ are one of the DDF operators in eqs.~\eqref{eq:ddf-ops}.\footnote{As we will see in Section~\ref{sec:checks}, $\mathcal{F}_r$ are the space-time modes, see eq.~\eqref{eq:modes-op}, of the vertex operators associated with the free bosons and free fermions, which were previously found in \cite{Gaberdiel:2021njm}. Therefore, an alternative perspective is to start with these vertex operators and \textit{derive} the DDF operators in eqs.~\eqref{eq:ddf-ops}.} Note that $\mathcal{F}_{k} \ket{\Omega_{P=0}}=0$ for $k\geq 0$. As an intermediate useful step, we claim that the vertex operator in eq.~\eqref{eq:generic-state-arg} satisfies\footnote{Since the correlation functions localize, one might alternatively take the form in eq.~\eqref{eq:arg-form} as an ansatz.}
\begin{equation} \label{eq:arg-form}
	(-1)^{f_A} V(G^-_{-1} \ket{\mathcal{A}};x,z) = \bar{\mathcal{A}}(z) \, \delta(\gamma(z)-x) \ ,
\end{equation}
for a local field $\bar{\mathcal{A}}(z)$. This relation holds up to BRST exact terms. What is important here is that the $x$-dependence comes only through the delta-function. This statement can be shown recursively. As the starting point, the $w=1$-twisted ground state in eq.~\eqref{eq:ground-state-pictures} obviously satisfies this relation, e.g.\ in picture $P=0$. Assume $V(\ket{\mathcal{A}};x,z)$ satisfies eq.~\eqref{eq:arg-form}. By applying the DDF operators in eqs.~\eqref{eq:ddf-ops}, one adds at most derivatives of $\gamma$ and other fields (but not $\beta$) to the vertex operator. This follows by considering OPEs of the following general form \cite{Witten:2012bh,Verlinde:1987sd}
\begin{equation} \label{eq:no-x-arg}
	\frac{F(t) G(z)}{\gamma(t)^k} \delta(\gamma(z)) = \frac{F(t) G(z)}{(\gamma(t)-\gamma(z))^k} \delta(\gamma(z)) \ ,
\end{equation}
for fields $F$ and $G$ on the world-sheet, and $k\geq 0$. Here we used that the delta-function imposes $\gamma(z)=0$, see \cite{Witten:2012bh}. Therefore, by taking the contour integral $t$ around $z$, only derivatives of $\gamma(z)$ appear in front of the delta-function. Since the only fundamental field that does not commute with $J^+_0=\beta_0$ is $\gamma(z)\rightarrow \gamma(z)-x$, see eq.~\eqref{eq:vertex-operator-x-z}, the field still takes the form given in eq.~\eqref{eq:arg-form}. Note that in the DDF operators in eqs.~\eqref{eq:ddf-ops}, $\beta$ does not appear.

How does eq.~\eqref{eq:arg-form} allow us to concretely define the space-time modes of a physical vertex operator? As we briefly review in Appendix~\ref{app:vertex}, there is a formal treatment of series including infinitely many terms with positive and negative powers, called `formal distributions'. In this formalism, one can for instance make sense of the following expression
\begin{equation} \label{eq:distribution-equality}
	\delta(z-x) = \sum_{n\in \mathbb{Z}} z^{n} x^{-n-1} \ .
\end{equation}
This is called the `formal delta-function', since as we discuss in Appendix~\ref{app:vertex}, this behaves as a delta-function. Roughly speaking, eq.~\eqref{eq:distribution-equality} holds because one can think of $\delta(z-x)$ as $1/(z-x)$ and consider its Taylor expansion. Although strictly speaking, the delta-function includes all the powers of $z$ and not only the positive or the negative powers, see eq.~\eqref{eq:delta-one-over}. We propose to interpret $\delta(\gamma(z)-x)$ as a formal distribution.\footnote{Note that this is then not a local field with respect to $\delta(\gamma(w))$ unless $x=0$, as there all infinitely many poles on the world-sheet, see eq.~\eqref{eq:no-x-arg}. However, as a distribution, each term is still a local field.} In fact, substituting the identity \eqref{eq:distribution-equality} into eq.~\eqref{eq:arg-form}, we recover the space-time mode expansion in eq.~\eqref{eq:mode-expansion} with
\begin{equation} \label{eq:vertex-mode}
	\mathcal{A}_r(z) = \bar{\mathcal{A}}(z) [\gamma(z)]^{r+h_A-1} \ , \quad r\in \mathbb{Z}-h_A \ . 
\end{equation}
Note that since the space-time free fermions are in the NS-sector, $x$ and $\gamma$ consistently have only integer powers. Having established eq.~\eqref{eq:mode-expansion}, we write the map $P$ in terms of space-time modes. Recall that $\mathcal{A}(x)$ in eq.~\eqref{eq:arg-ax} maps physical states to physical states for any value of $x$. Therefore, by taking residue of both sides of eq.~\eqref{eq:arg-ax} after multiplying by an appropriate power of $x$, we see that
\begin{equation} \label{eq:modes-op}
	\mathcal{A}_r = \oint dz \, \mathcal{A}_r(z)
\end{equation}
also maps physical states to physical states.  In fact, we have
\begin{equation} \label{eq:space-time-mode-expansion}
	\mathcal{A}(x-y) = W(\ket{\mathcal{A}},x-y) = \sum_{r\in\mathbb{Z}-h_A} \mathcal{A}_r \, (x-y)^{-r-h_A} \ ,
\end{equation}
where both sides act on the world-sheet Fock space. Inserting the space-time mode expansion in eq.~\eqref{eq:space-time-mode-expansion} into eq.~\eqref{eq:p-product-arg}, we get
\begin{equation}\label{eq:ope-mode}
	P[\mathcal{A},\mathcal{B}](x,y;w) = \sum_{r\in\mathbb{Z}-h_A} V(\mathcal{A}_r \ket{\mathcal{B}};y;w) \, (x-y)^{-r-h_A} \ .
\end{equation}
In the rest of this section, we show how eq.~\eqref{eq:ope-mode} allows us to argue that the space-time OPEs are correctly reproduced from the world-sheet. The idea is to define an isomorphism $\phi$ from the space-time CFT to the set of physical states such that it preserves the OPE structure. Let us explain this in detail. We denote the Fock space of the untwisted part of the space-time theory by $\mathcal{V}_{\text{s.}}$, and define $\mathcal{V}_{\text{w.}}$ as the subspace of the $w=1$ spectrally flowed physical states modulo BRST exactness and picture ambiguity. In the space-time theory, we denote the vertex operator map as $Y$. It satisfies the `duality' relation of \cite{Gaberdiel:1998fs,Goddard:1989dp}
\begin{equation} \label{eq:duality-st}
	Y(\ket{S},x) Y(\ket{T},y) = Y(Y(\ket{S},x-y)\ket{T},y) \ .
\end{equation}
On the world-sheet, the vertex operator map $W$ is defined in eq.~\eqref{eq:arg-ax} as
\begin{equation}
	W(\ket{\mathcal{A}},x)=(-1)^{f_A} V(G^-_{-1}\ket{\mathcal{A}};x,z)_0 \ ,
\end{equation}
where the zero mode is on the world-sheet. Note that as we discussed, $W(\ket{\mathcal{A}},x)$ acts on $\mathcal{V}_{\text{w.}}$ and the OPE prescription in eq.~\eqref{eq:p-product-arg} is how one calculates OPEs on the right-hand side of eq.~\eqref{eq:duality-st}. In particular, we have\footnote{The duality relation also holds for $W$ in terms of (anti-)commutator of world-sheet vertex operators. As we mentioned, the physical vertex operators are generically non-local fields. Therefore, the duality relation is not directly useful for our purpose and we define the OPE as in eq.~\eqref{eq:w-duality-relation}, also see eq.~\eqref{eq:app-exchange} and below.}
\begin{equation} \label{eq:w-duality-relation}
	(-1)^{f_{P[\mathcal{S},\mathcal{T}]}} (G^-_{-1} P[\mathcal{S},\mathcal{T}](x,y;w))_0 = W(W(\ket{\mathcal{S}},x-y)\ket{\mathcal{T}},y) \ .
\end{equation}
Motivated by this observation, we define a map $\phi:\mathcal{V}_{\text{s.}} \rightarrow \mathcal{V}_{\text{w.}}$ by
\begin{equation} \label{eq:def-phi}
	\phi(\ket{A}) = \ket{\mathcal{A}} \ , \quad \bar{\phi}(Y(\ket{A},x))=W(\phi(\ket{A}),x) \ ,
\end{equation}
and extend it linearly to $\mathcal{V}_{\text{s.}}$. Note that the map $\bar{\phi}$ acts on the vertex operators and is chosen to be consistent with $\phi$. The key idea is the following: if we show
\begin{equation} \label{eq:argue-to-show}
	\bar{\phi}(Y(\ket{S},x)) \phi(\ket{T}) = \phi(Y(\ket{S},x)\ket{T}) \ ,
\end{equation}
then it follows that the OPEs in eq.~\eqref{eq:ope-mode} are identical to the space-time OPEs. This is because the OPE is calculated as follows
\begin{align}
	W(W(\ket{\mathcal{S}},x-y)\ket{\mathcal{T}},y) &= W(\bar{\phi}(Y(\ket{S},x-y)) \phi(\ket{T}),y) \\&= W(\phi(Y(\ket{S},x-y)\ket{T}),y) = \bar{\phi}(Y(Y(\ket{S},x-y)\ket{T},y)) \ . \nonumber
\end{align}
Here in the first and last equalities we used eq.~\eqref{eq:def-phi}, and in the second equality we used eq.~\eqref{eq:argue-to-show}. Therefore, our aim is to show eq.~\eqref{eq:argue-to-show} which subsequently implies that eq.~\eqref{eq:p-product-arg} agrees with the space-time OPEs. In fact, considering the mode expansion in eq.~\eqref{eq:mode-expansion}, we see that eq.~\eqref{eq:argue-to-show} is satisfied\- if
\begin{equation} \label{eq:arg-mode-to-show}
	\phi(S_r \ket{T})=\mathcal{S}_r \ket{\mathcal{T}} \ , \quad r\in\mathbb{Z}-h_S \ .
\end{equation}
In the remainder of this section, we show eq.~\eqref{eq:arg-mode-to-show}.

Let us be more precise about how the map $\phi$ is defined. As we mentioned before, any state (with zero momenta and winding numbers) in the untwisted sector of the space-time theory can be uniquely written as a linear sum of the following states
\begin{equation} \label{eq:st-generic-state}
	\ket{A}=F^{(1)}_{-s_1} \cdots F^{(n)}_{-s_n} \ket{0} \ ,
\end{equation}
where $F^{(j)}$ are the free bosons and fermions, see Appendix~\ref{app:t4}. We define
\begin{equation} \label{eq:def-phi-generic-state}
	\phi(\ket{A}) = \ket{\mathcal{A}} = \mathcal{F}^{(1)}_{-s_1} \cdots \mathcal{F}^{(n)}_{-s_n} \ket{\Omega_{P=0}} \ ,
\end{equation}
and extend it linearly to $\mathcal{V}_{\text{s.}}$, see eq.~\eqref{eq:ground-state-pictures}. In fact, the map $\phi$ is an isomorphism, which essentially follows from the analyses of \cite{Eberhardt:2018ouy,Naderi:2022bus}.\footnote{If $\ket{A}$ has non-zero momenta and winding numbers, there is a natural generalization of this map, see footnote~8 in \cite{Naderi:2022bus}.} Our strategy for proving eq.~\eqref{eq:arg-mode-to-show} is simple. If $\ket{T}=\ket{0}$, it is necessary that the actions on the `vacuums' are isomorphic
\begin{equation} \label{eq:locality-assumption}
	\phi(S_r\ket{0}) = \mathcal{S}_{r} \ket{\Omega_{P=0}} \ , \quad r\in\mathbb{Z}-h_S \ .
\end{equation}
Moreover, by writing the state $\ket{T}$ as in eq.~\eqref{eq:st-generic-state}, we see that if further
\begin{equation} \label{eq:s-f-modes}
	\bar{\phi}([S_r,F_s])=[\mathcal{S}_r,\mathcal{F}_s]
\end{equation}
holds, eq.~\eqref{eq:arg-mode-to-show} is satisfied. This statement is explained in more detail in Appendix~\ref{app:ope}, together with an explicit example that is discussed in Appendix~\ref{app:explicit-arg}.

In order to show eq.~\eqref{eq:locality-assumption} concretely, we start with the case where $r\geq -h_S$. We use eqs.~\eqref{eq:vertex-mode} and \eqref{eq:ground-state-pictures} to directly calculate the right-hand side of eq.~\eqref{eq:locality-assumption} in a specific picture
\begin{equation}
	\oint_{C_w} dz \, \bar{\mathcal{S}}(z) \gamma^n(z) \partial \gamma(w) \delta(\gamma(w)) e^{i\sigma} =(-1)^{f_A} e^{i\sigma}\delta_{n,-1} \bar{\mathcal{S}}(w) \delta(\gamma(w)) \ , \quad (n\geq -1) \ .
\end{equation}
Here we have used that inside $\bar{\mathcal{S}}(z)$ there is no $\beta$, see the discussion below eq.~\eqref{eq:arg-form}. In particular, when $n\geq 0$, there is no pole but when $n=-1$, similar to eq.~\eqref{eq:no-x-arg}, there is a simple pole and we get the right-hand side. Note that we have exchanged $e^{i\sigma}$ with $\bar{\mathcal{S}}$. This shows that eq.~\eqref{eq:locality-assumption} holds up to BRST exact terms for $r\geq -h_S$, see eq.~\eqref{eq:arg-form} and Appendix~\ref{app:ope}. For $r<-h_S$, as we discuss in Appendix~\ref{app:ope}, eq.~\eqref{eq:locality-assumption} holds, and it follows essentially from the definition of the $x$-basis in eq.~\eqref{eq:vertex-operator-x-z}. This can be equivalently written as
\begin{equation} \label{eq:def-vertex-st-arg}
	\phi(e^{x T_{-1}} S_{-h_S}\ket{0}) = e^{x\mathcal{L}_{-1}} \mathcal{S}_{-h_S} \ket{\Omega_{P=0}} \ ,
\end{equation}
where $T_{-1}$ is the translation operator in the space-time and $\mathcal{L}_{-1}=J^+_0$.

Now we show eq.~\eqref{eq:s-f-modes}. Let us collectively denote $\mathcal{F}$ as the vertex operators associated with the free bosons and free fermions, see eqs.~\eqref{eq:fermions} and \eqref{eq:vertex-bosons}. As one could guess, and we will show in Section~\ref{sec:checks}, the space-time modes in eq.~\eqref{eq:modes-op} associated with $\mathcal{F}$ agree with the DDF operators in eqs.~\eqref{eq:ddf-ops}, see eqs.~\eqref{eq:fermions-modes-4}, \eqref{eq:barred-bosons-modes-4} and \eqref{eq:unbarred-bosons-modes-4} and the discussion below. Therefore, they satisfy the correct space-time algebra given in eq.~\eqref{eq:ddf-algebra}. Given two physical vertex operators $\mathcal{S}$ and $\mathcal{T}$, we want to show
\begin{equation} \label{eq:anti-comm-s-t}
	[\mathcal{S}_n,\mathcal{T}_m] = \oint_{C_0} dy \, y^{m+h_T-1} \oint_{C_y} dx \, x^{n+h_S-1} W(W(\ket{\mathcal{S}},x-y)\ket{\mathcal{T}},y) \ ,
\end{equation}
where the bracket denotes the super-commutator on the world-sheet. To show this, we use eqs.~\eqref{eq:w-duality-relation} and \eqref{eq:explicit-p-formula} to write
\begin{equation} \label{eq:anti-comm-w}
	(G^-_{-1} P[\mathcal{S},\mathcal{T}](x,y;w))_0 = \oint_{C_0} dw \, \oint_{C_w} dz \, V(G^-_{-1}\ket{\mathcal{S}};x,z) V(G^-_{-1}\ket{\mathcal{T}};y,w) \ .
\end{equation}
Here we have exchanged $G^-_{-1}$ with $W(\ket{\mathcal{A}};x-y)$. Using eq.~\eqref{eq:mode-expansion}, we obtain
\begin{equation} \label{eq:residue-arg}
	[\mathcal{S}_n,\mathcal{T}_m] = \text{Res}_y \, y^{m+h_T-1} \text{Res}_x \, x^{n+h_S-1} (-1)^{f_S+f_T} (G^-_{-1}P[\mathcal{S},\mathcal{T}](x,y;w))_0 \ ,
\end{equation}
where $f_{P[\mathcal{S},\mathcal{T}]}=f_S+f_T$. In Appendix~\ref{app:ope}, we show
\begin{equation} \label{eq:res-contour}
	(-1)^{f_{S}+f_T} \text{Res}_x \, x^{l} (G^-_{-1}P[\mathcal{S},\mathcal{T}](x,y;w))_0 = \oint_{C_y} dx \, x^{l} \, W(W(\ket{\mathcal{S}},x-y)\ket{\mathcal{T}},y) \ ,
\end{equation}
for an integer $l$. This implies that eq.~\eqref{eq:anti-comm-s-t} holds. Let us now consider the space-time OPE $P[\mathcal{F},\mathcal{S}](x,y;w)$, where $\mathcal{F}$ is the vertex operator of free bosons and free fermions, and $\mathcal{S}$ is of the form in eq.~\eqref{eq:def-phi-generic-state}. As in this case eqs.~\eqref{eq:locality-assumption} and \eqref{eq:s-f-modes} are satisfied, see eqs.~\eqref{eq:ddf-algebra} and \eqref{eq:ddf-identity-operator}, it is then straightforward to show
\begin{equation}
	W(W(\ket{\mathcal{F}},x-y)\ket{\mathcal{S}},y) = \bar{\phi}(Y(Y(\ket{F},x-y)\ket{S},y)) \ .
\end{equation}
Substituting this back in eq.~\eqref{eq:anti-comm-s-t}, we see that eq.~\eqref{eq:s-f-modes} is satisfied. Therefore, eqs.~\eqref{eq:locality-assumption} and \eqref{eq:s-f-modes} hold. As we discussed, we conclude that the OPEs in eq.~\eqref{eq:p-product-arg} agree with the space-time OPEs.
%%%%%%%%%%%%%%%%%%%%%%%%%%%%%%%%%%%%%%%%%%%%%%%%%%%%%%%%%%%%
\section{Consistency checks and examples} \label{sec:checks}
In this section, we illustrate the OPE calculations directly on the world-sheet: we use the map $P$ defined in the previous section to calculate the space-time OPEs. In particular, we treat the delta-function $\delta(\gamma(z)-x)$ as $1/(x-\gamma(z))$ assuming $|x|>|\gamma(z)|$, see eq.~\eqref{eq:delta-one-over}. This is then equivalent to considering the `singular' part of eq.~\eqref{eq:distribution-equality}. Therefore, here we are only able to reproduce the singular part of the OPEs. In Appendix~\ref{app:example}, we discuss two more non-trivial examples of OPE calculations using the method discussed in this section: we calculate the OPE of two supercurrents and the OPE of the stress-tensor with itself, and both computations agree with the space-time result.
%%%%%%%%%%%%%%%%%%%%%%%%%%%%%%%%%%%%%%%%%%%%%%%
\subsection{Free fermions} \label{sec:fermions}
The vertex operators associated with the space-time free fermions written in the variables of Section~\ref{sec:review} are \cite{Gaberdiel:2021njm}
\begin{equation} \label{eq:fermions}
	\Psi^{+,j}(x,z) = -e^{if_1} \delta(\gamma(z)-x) e^{\rho+i\sigma+iH^j} \ , \quad \Psi^{-,j}(x,z) = e^{-if_2} \delta(\gamma(z)-x) e^{\rho+i\sigma+iH^j} \ ,
\end{equation}
where $j\in\{1,2\}$. Before we begin the calculation of their OPEs, let us calculate their space-time modes from eq.~\eqref{eq:vertex-mode}, which gives
\begin{equation} \label{eq:fermions-modes-4}
	\Psi^{+,j}_r(z)=e^{if_1} e^{\rho+iH^j} \gamma^{r-\frac{1}{2}} \ , \quad \Psi^{-,j}_r(z)=-e^{-if_2} e^{\rho+iH^j} \gamma^{r-\frac{1}{2}} \ .
\end{equation}
Note that the additional minus sign comes because $f_{\Psi^{\alpha,j}}=1 \bmod 2$, see Appendix~\ref{app:ope}. This upon integration agrees with the DDF operators in eq.~\eqref{eq:ddf-fermions}. Since the DDF operators satisfy the correct space-time algebra in eq.~\eqref{eq:ddf-algebra}, the modes also satisfy the expected algebra and this completes our claim in Section~\ref{sec:argument} for the free fermions.

Now we would like to calculate the following OPE
\begin{equation}
	P[\Psi^{\alpha,j},\Psi^{\beta,k}](x,y;w) \ ,
\end{equation}
where $\alpha,\beta\in \{+,-\}$, see eq.~\eqref{eq:p-product-arg}. More explicitly, using eq.~\eqref{eq:explicit-p-formula} this equals
\begin{equation}
	\oint_{C_w} dz \, [O(x,z;y,w)] \ , \quad O(x,z;y,w)= -[G^-_{-1} \Psi^{\alpha,j}(x,z)][\Psi^{\beta,k}(y,w)] \ .
\end{equation}
As we mentioned before, in calculating the OPEs, we describe the $(\beta,\gamma)$ fields in the path-integral formalism, while the rest are honest vertex operators in the CFT. The only non-zero contribution comes from the case $\alpha\neq \beta$ and $j\neq k$, and it equals\footnote{We are assuming $|x|>|y|$ and in particular, $x\neq y$. The symbol `$\sim$' denotes the equality up to singular terms, both on the world-sheet and in the space-time.}
\begin{equation}
	O(x,z;y,w) \sim -\frac{\epsilon^{\alpha\beta} \epsilon^{kj}}{z-w} e^{if_1-if_2} e^{2\rho+i\sigma+iH} \delta(\gamma(z)-x) \delta(\gamma(w)-y) \ . 
\end{equation}
Taking the integral over $z$ gives
\begin{equation} \label{eq:ope-fermions-1}
	P[\Psi^{\alpha,j},\Psi^{\beta,k}](x,y;w)\sim-\epsilon^{\alpha\beta} \epsilon^{kj} e^{if_1-if_2} e^{2\rho+i\sigma+iH} \delta(\gamma(w)-x) \delta(\gamma(w)-y) \ . 
\end{equation}
In order to relate this expression to a quantity in the space-time, we use that the second delta-function $\delta(\gamma(w)-y)$ imposes $\gamma(w)=y$, also see \cite{Kutasov:1999xu} and Appendix~\ref{app:vertex}. We therefore get
\begin{align} \label{eq:delta-impose}
	\delta(\gamma(w)-x)\delta(\gamma(w)-y)&=\delta(y-x) \delta(\gamma(w)-y)\\&= \delta(x-y)\delta(\gamma(w)-y)= \frac{\delta(\gamma(w)-y)}{x-y} \ , \quad (|x|>|y|) \ . \nonumber
\end{align}
In the second line, we first used that $\delta(x-y)=\delta(y-x)$, see Appendix~\ref{app:vertex}. In the next equality, we used eq.~\eqref{eq:delta-one-over}, i.e.\ we treated $\delta(x-y)$ as $1/(x-y)$ for $|x|>|y|$. Using this, we note that the expression in eq.~\eqref{eq:ope-fermions-1} is proportional to the $w=1$-twisted ground state in eq.~\eqref{eq:ground-state-pictures} in picture $P=-2$. Simplifying the expression gives
\begin{equation}
	P[\Psi^{\alpha,j},\Psi^{\beta,k}](x,y;w) \sim \frac{\epsilon^{\alpha\beta} \epsilon^{kj} \Omega_{P=-2}(y,w)}{(x-y)} \ ,
\end{equation}
which is the correct space-time result, see Appendix~\ref{app:t4}.
%%%%%%%%%%%%%%%%%%%%%%%%%%
\subsection{Free bosons} \label{sec:bosons}
The fields that correspond to the free bosons are \cite{Gaberdiel:2021njm}
\begin{subequations} \label{eq:vertex-bosons}
\begin{equation} \label{eq:barred}
	\partial \bar{\mathcal{X}}^j(x,z) = \delta(\gamma(z)-x) e^{i\sigma} \partial \bar{X}^j \ ,
\end{equation}
\begin{equation}
	\partial \mathcal{X}^j(x,z) = e^{2if_1-2if_2} (\partial \gamma(z))^{-2}\delta(\gamma(z)-x) e^{4\rho+i\sigma+2iH} \partial X^j \ .
\end{equation}
\end{subequations}
Let us first identify their modes using eq.~\eqref{eq:vertex-mode}. For the barred bosons we claim that
\begin{equation} \label{eq:barred-bosons-modes-4}
	\partial \bar{\mathcal{X}}^j_n(z) = \partial \bar{X}^j(z) \gamma^n \ . 
\end{equation}
Upon integration, this agrees with the DDF operators of the barred bosons, see eq.~\eqref{eq:ddf-bosons}. For the unbarred boson, we get instead
\begin{equation} \label{eq:unbarred-bosons-modes-4}
	\partial \mathcal{X}^j_n(z) = e^{2if_1-2if_2} [\partial \gamma(z)]^{-2} e^{4\rho+2iH} \partial X^j \gamma^{n} \ .
\end{equation}
In fact, by applying the picture-raising operator $P_+$ twice, see eq.~\eqref{eq:picture-raising}, this upon integ\-ration agrees with the DDF operators in eq.~\eqref{eq:ddf-bosons}.\footnote{The difference between $(P_+)^2 [\partial \mathcal{X}^j_n]$ and the corresponding integrand in eq.~\eqref{eq:ddf-bosons} is a total derivative.} As the DDF operators satisfy the expected space-time algebra, see eq.~\eqref{eq:ddf-algebra}, the modes also satisfy the correct algebra and this shows our claim in Section~\ref{sec:argument} for the free bosons.

Now we will calculate the following OPE
\begin{equation}
	P[\partial \bar{\mathcal{X}}^j,\partial \mathcal{X}^k](x,y;w) \ .
\end{equation}
Similar to the case of the free fermions, using eq.~\eqref{eq:explicit-p-formula} this equals
\begin{equation}
	\oint_{C_w} dz \, [O(x,z;y,w)] \ , \quad O(x,z;y,w)= [G^-_{-1} \partial \bar{\mathcal{X}}^j(x,z)][\partial \mathcal{X}^k(y,w)] \ .
\end{equation}
Calculating the world-sheet OPE gives
\begin{equation}
	O(x,z;y,w) \sim \frac{\delta^{jk}}{(z-w)^2} [\partial \gamma(w)]^{-2} e^{2if_1(w)-2if_2(w)} e^{(4\rho+i\sigma+2iH)(w)} \delta(\gamma(z)-x) \delta(\gamma(w)-y) \ , 
\end{equation}
thus leading to
\begin{equation} \label{eq:bosons-ope-1}
	P[\partial \bar{\mathcal{X}}^j,\partial \mathcal{X}^k](x,y;w) \sim \delta^{jk}[\partial_w \delta(\gamma(w)-x)] (\partial\gamma)^{-2} e^{2if_1-2if_2} e^{4\rho+i\sigma+2iH} \delta(\gamma(w)-y) \ .
\end{equation}
We use the chain rule to write
\begin{equation} \label{eq:chain-rule}
	\partial_w \delta(\gamma(w)-x)=-[\partial\gamma(w)] \partial_x\delta(\gamma(w)-x) = -[\partial \gamma(w)] \partial_x \delta(x-\gamma(w)) = \frac{\partial \gamma(w)}{(x-\gamma(w))^2} \ .
\end{equation}
In the first equality, we used the chain rule and in the second equality we used the fact that $\delta(a-b)=\delta(b-a)$, see Appendix~\ref{app:vertex}. In the last equality, we treated $\delta(x-a)\sim 1/(x-a)$ for $|x|>|a|$, see eq.~\eqref{eq:delta-one-over}.
Making use of the appearance of the second delta-function in eq.~\eqref{eq:bosons-ope-1}, we have
\begin{equation}
	P[\partial \bar{\mathcal{X}}^j,\partial \mathcal{X}^k](x,y;w) \sim \frac{\delta^{jk} e^{2if_1-2if_2} (\partial \gamma)^{-1} e^{4\rho+i\sigma+2iH} \delta(\gamma(w)-y)}{(x-y)^2} \ .
\end{equation}
Finally, we use eq.~\eqref{eq:ground-state-pictures} to write
\begin{equation}
	P[\partial \bar{\mathcal{X}}^j,\partial \mathcal{X}^k](x,y;w) \sim \frac{\delta^{jk} \Omega_{P=-4}(y,w)}{(x-y)^2} \ ,
\end{equation}
which agrees with the space-time OPE, see Appendix~\ref{app:t4}.
%%%%%%%%%%%%%%%%%%%%%%%%%%%%%%%%%%%%%%%%%%%%%%%%%%%%%%%
\subsection{The identity operator} \label{sec:identity}
The vertex operator of the $w=1$-twisted ground state from eq.~\eqref{eq:ground-state-pictures} in picture $P=0$ is
\begin{equation}
	\Omega(x,z)=(\partial \gamma) \delta(\gamma(z)-x) e^{i\sigma}\ .
\end{equation}
Here for brevity, we drop the subscript of the picture number. Using eq.~\eqref{eq:vertex-mode}, we read the space-time modes
\begin{equation}
	\Omega_r(z) = \gamma^{r-1} (\partial \gamma) \ .
\end{equation}
Therefore, using the definition in eq.~\eqref{eq:modes-op} we get
\begin{equation}
	\Omega_r = \oint dz \, \gamma^{r-1} (\partial \gamma) \ .
\end{equation}
If $r\neq 0$, the integrand is a total derivative and therefore vanishes \cite{Giveon:1998ns,Eberhardt:2019qcl}. If $r=0$, we see that $\Omega_0$ agrees with the DDF operator of the identity $\mathcal{I}$ in eq.~\eqref{eq:ddf-identity-operator}. This matches with the space-time expectation.\footnote{One can define a normal-ordered product using the usual integral definition of the radial normal-ordering and the map $P$ in eq.~\eqref{eq:p-product-arg}, see e.g.\ \cite{DiFrancesco:1997nk}. This integral can be evaluated because of the appearance of delta-functions. As a sanity check, we have checked that the normal-ordering of $\Omega$ with itself, and the fermions with themselves reproduce the correct space-time result.}
%%%%%%%%%%%%%%%%%%%%%%%%%%%%%%%%%%%%%%%%%%%%%%
\section{\texorpdfstring{$\mathcal{W}_{\infty}$}{Winfinity} algebra from the world-sheet} \label{sec:w-algebra}
In this section, we perform another sanity check for the proposal in eq.~\eqref{eq:p-product-arg}. In particular, we consider the $\mathcal{W}$-algebra that is realized in the untwisted sector of the symmetric orbifold of $\mathbb{T}^4$. As the topic is not directly related to the rest of the paper, we begin by reviewing some facts about higher spin theories and their relation to $\mathcal{W}_{\infty}$ algebras, see \cite{Gaberdiel:2012uj} for an extensive review. A familiar reader may directly continue with eq.~\eqref{eq:winfinity-sup}.

It is possible to consider consistent gauge theories on $\text{AdS}_3$ with a finite number of spins $s=2,3,\cdots,N$, where $N$ is an integer \cite{Vasiliev:1995dn,Campoleoni:2010zq,Henneaux:2010xg}. These theories are described by Chern-Simon theories on $\mathfrak{sl}(N) \times \mathfrak{sl}(N)$ \cite{Achucarro:1986uwr,Witten:1988hc}, see also \cite{Collier:2023fwi,Collier:2024mgv} for a recent development in the case of gravity. Similar to this, it is possible to consider a theory on $\text{AdS}_3$ with infinitely many higher spin fields. These theories are described by Chern-Simons theories based on `higher spin algebras' $\mathfrak{hs}[\mu]\times \mathfrak{hs}[\mu]$ \cite{Vasiliev:1995dn,Vasiliev:1999ba,Prokushkin:1998bq,Prokushkin:1998vn}, where $\mu$ is a free parameter which for $\mu=N$, $\mathfrak{hs}[N]$ truncates to $\mathfrak{sl}(N)$. In \cite{Gaberdiel:2010pz}, it was argued that this higher spin theory\footnote{Strictly speaking, there are also two complex scalars, see \cite{Gaberdiel:2010pz}.} is dual to a `t Hooft limit of what are called the $\mathcal{W}_{N,k}$ minimal models
\begin{equation} \label{eq:minimal-model}
	\frac{\text{SU}(N)_{k} \times \text{SU}(N)_1}{\text{SU}(N)_{k+1}} \ .
\end{equation}
Here $N$ and $k$ are taken to be large while the `t Hooft coupling $\lambda=\frac{N}{N+k}$ is kept fixed. This coset CFT realizes a $\mathcal{W}_N$ algebra, which can be seen as a special case of a $\mathcal{W}_{\infty}$ algebra \cite{Pope:1989ew,Pope:1991ig,Gaberdiel:2011wb}. These algebras, which we denote by $\mathcal{W}_{\infty}[\nu]$, depend on two parameters: the central charge $c$ and a free parameter $\nu$ that enters the commutation relations, see \cite{Pope:1989ew,Pope:1989sr,Pope:1991ig,Gaberdiel:2012uj}. For $\nu=N$, this algebra truncates to $\mathcal{W}_N$ which has one generator for each integer spin $2\leq s\leq N$, see \cite{Linshaw:2017tvv,Gaberdiel:2012ku,Eberhardt:2019xmf}. On the other hand, the higher spin theory, after an appropriate `quantization' of the commutation relations, realizes $\mathcal{W}_{\infty}[\mu]$ via asymptotic symmetries \cite{Brown:1986nw,Henneaux:2010xg,Campoleoni:2010zq,Gaberdiel:2012uj}. In fact, using the `triality relation' of \cite{Gaberdiel:2012ku}, one can show that these two $\mathcal{W}$-algebras agree, and in particular, in the `t Hooft limit, they both realize a $\mathcal{W}_{\infty}[\mu]$ algebra where $\mu=\lambda$. This duality is referred to as `minimal model holography'.

Tensionless strings on $\text{AdS}_3$ are in fact related to a similar duality that we just discussed, see \cite{Gaberdiel:2013vva,Gaberdiel:2014cha,Gaberdiel:2011zw,Gaberdiel:2014yla,Eberhardt:2018plx} for the precise relation. Gaberdiel and Gopakumar, by studying a higher spin theory in \cite{Gaberdiel:2013vva} based on the Lie algebra $D(2,1|\alpha)$ and taking a large level limit in \cite{Gaberdiel:2014cha}, were able to argue that the (perturbative part of the) resulting higher spin theory\- describes a subsector of the full string theory. In the space-time, this corresponds to considering (a subsector of) the `untwisted' sector of the symmetric orbifold of $\mathbb{T}^4$. In fact, the authors\- showed in \cite{Gaberdiel:2015mra} that the symmetric orbifold of $\mathbb{T}^4$ contains two distinct $\mathcal{W}_{1+\infty}[0]$ and $\mathcal{W}_{\infty}[1]$ algebras which together generate what they called the `higher spin square' symmetry\- algebra\-. In particular\-, the space-time theory contains a supersymmetric $s\mathcal{W}^{(\mathcal{N}=4)}_{\infty}[0]$ algebra\- \cite{Pope:1991ig,Depireux:1990df,Bergshoeff:1990yd}, see \cite{Gaberdiel:2015mra,Gaberdiel:2013vva,Gaberdiel:2014cha}. In the remainder of this section, we will discuss how the (anti-)commutation relations of this algebra are satisfied on the world-sheet.

To begin with, we note that the spin content of the $s\mathcal{W}^{(\mathcal{N}=4)}_{\infty}[0]$ algebra can be re\-arranged as \cite{Gaberdiel:2013vva,Gaberdiel:2015uca}
\begin{equation} \label{eq:winfinity-sup}
	s\mathcal{W}^{(\mathcal{N}=4)}_{\infty}[0] : (\mathcal{N}=4) \oplus \bigoplus_{s=1}^{\infty} R^{(s)} \ .
\end{equation}
Here $(\mathcal{N}=4)$ denotes the small $\mathcal{N}=4$ algebra of $\mathbb{T}^4$, see Appendix~\ref{app:t4}, and $R^{(s)}$ are the supermultiplets with spin $s\geq 1$. In fact, $R^{(1)}$ contains a spin-$1$ field which is not part of the small $\mathcal{N}=4$ algebra. One expects that together with the $\mathcal{N}=4$ algebra they generate the whole $s\mathcal{W}^{(\mathcal{N}=4)}_{\infty}[0]$, see e.g.\ \cite{Gaberdiel:2013vva,Gaberdiel:2015uca}. For this reason and also for simplicity, here we only explicitly discuss this low-lying spin-$1$ field in $R^{(1)}$ and denote it in the space-time by $K$, also see the discussion in the Conclusions and remarks. We will reproduce the space-time commutation relations of $K$ with the $\mathcal{N}=4$ algebra from the world-sheet perspective. Together with the argument given in Section~\ref{sec:argument}, it is then clear how the $s\mathcal{W}^{(\mathcal{N}=4)}_{\infty}[0]$ algebra is realized on the world-sheet. In \cite{Gaberdiel:2015uca}, it is discussed that
\begin{equation} \label{eq:space-time-k}
	K = -\frac{1}{2}\Big[ (\psi^{+,1} \psi^{-,2})+(\psi^{+,2} \psi^{-,1})\Big] \ ,
\end{equation}
see Appendix~\ref{app:t4}. Using the DDF operators in eq.~\eqref{eq:ddf-fermions}, the corresponding vertex operator on the world-sheet is
\begin{equation} \label{eq:spin1}
	\mathcal{K}_{P=-2}(x,z)=-\frac{1}{2} e^{if_1-if_2} (\partial \gamma)^{-1} \delta(\gamma(z)-x) e^{2\rho+i\sigma+iH} \partial(iH^1-iH^2) \ .
\end{equation}
By applying the picture-raising operator $P_+$ in eq.~\eqref{eq:picture-raising}, we get
\begin{equation} \label{eq:spin1-p=0}
	\mathcal{K}_{P=0}(x,z)=\frac{1}{2} \delta(\gamma(z)-x) e^{i\sigma} \partial(iH^1-iH^2) \ .
\end{equation}
Let us now calculate for instance the OPE of $\mathcal{K}_{P=0}$ with itself. This is very similar to the calculation given in Section~\ref{sec:bosons} and we get
\begin{equation}
	P[\mathcal{K}_{P=0},\mathcal{K}_{P=0}](x,y;w) \sim \frac{(\partial\gamma) e^{i\sigma}\delta(\gamma(w)-y)}{2(x-y)^2} = \frac{\Omega_{P=0}(y,w)}{2(x-y)^2} \ ,
\end{equation}
where we have used eq.~\eqref{eq:ground-state-pictures}. This agrees with the space-time result, see Appendix~\ref{app:t4}. We continue with studying the OPEs of $\mathcal{K}$ with the $\mathcal{N}=4$ superconformal generators in Appendix~\ref{app:t4} from the world-sheet. To begin with, we consider the OPEs of $\mathcal{K}$ with the R-symmetry currents $\mathcal{J}^a$, where the associated vertex operators are \cite{Gaberdiel:2021njm}
\begin{equation}
	\mathcal{J}_{P=0}^a = K^a \delta(\gamma(z)-x) e^{i\sigma} \ .
\end{equation}
It is clear that their OPE with $\mathcal{K}_{P=0}$ in eq.~\eqref{eq:spin1-p=0} is trivial, which is indeed the case in the space-time theory, see eq.~\eqref{eq:space-time-k} and Appendix~\ref{app:t4}. Using this, if we want to calculate the OPE of $\mathcal{K}$ with the stress-tensor, it is enough to consider its OPE with the topologically twisted stress-tensor $\mathcal{T}$ in eq.~\eqref{eq:topologically-twisted-state}. In fact, using similar methods discussed in Section~\ref{sec:checks} and Appendix~\ref{app:example}, it is straightforward to show
\begin{equation}
	P[\mathcal{K}_{P=0},\mathcal{T}_{P=-4}](x,y;w) \sim \frac{e^{2if_1-2if_2} (\partial \gamma)^{-2} \delta(\gamma(w)-y) e^{4\rho+i\sigma+2iH} \partial(iH^1-iH^2)}{2(x-y)^2} \ .
\end{equation}
This field is exactly the field associated with $K$ in picture $P=-4$
\begin{equation}
	P[\mathcal{K}_{P=0},\mathcal{T}_{P=-4}](x,y;w) \sim \frac{\mathcal{K}_{P=-4}(y,w)}{(x-y)^2} \ .
\end{equation}
This is the expected space-time OPE. The calculation of the OPEs with the supercurrents is similar, and they only have a simple pole with $K$. Let us illustrate this with an example. The space-time OPE of $K$ in eq.~\eqref{eq:space-time-k} and $G^+$ in eq.~\eqref{eq:app-t4-gp} is
\begin{equation} \label{eq:k-gp-space-time}
	K(x) G^+(y) \sim \frac{G^{+,\prime}(y)}{2(x-y)} \ , \quad G^{+,\prime} = \partial \bar{X}^1 \psi^{+,1} - \partial \bar{X}^2 \psi^{+,2} \ .
\end{equation}
Now we perform this calculation on the world-sheet. The physical vertex operator associated to the supercurrent $G^+$ can be written as\footnote{This field is related by the picture-raising operator $P_+$ defined in eq.~\eqref{eq:picture-raising} to the field in eq.~\eqref{eq:app-examples-gp}.}
\begin{equation}
	\mathcal{G}^+_{P=-1}(x,z)=-e^{if_1} (\partial \gamma)^{-1} \delta(\gamma(z)-x) e^{\rho+i\sigma} G^+_C \ .
\end{equation}
By directly calculating the OPE in eq.~\eqref{eq:p-product-arg}, we get
\begin{equation} \label{eq:k-gp}
	P[\mathcal{K}_{P=0},\mathcal{G}^+_{P=-1}](x,y;w) \sim -\frac{e^{if_1} (\partial \gamma)^{-1} \delta(\gamma(w)-y) e^{\rho+i\sigma} G^{+,\prime}_C}{2(x-y)} \ ,
\end{equation}
where we have defined
\begin{equation}
	G^{+,\prime}_C = \partial \bar{X}^1 e^{iH^1} - \partial \bar{X}^2 e^{iH^2} \ .
\end{equation}
The claim is that the field appearing in eq.~\eqref{eq:k-gp} is the physical vertex operator associated to $\tfrac{1}{2} G^{+,\prime}$, see eq.~\eqref{eq:k-gp-space-time}. This can be directly checked using the DDF operators in eqs.~\eqref{eq:ddf-ops}, and therefore, the world-sheet OPE agrees with the space-time OPE. The same argument holds as well for the OPEs of $\mathcal{K}$ with the other supercurrents in Appendix~\ref{app:t4}, and they all agree with the space-time result.
%%%%%%%%%%%%%%%%%%%%%%%%%%%%%%%%%%%%%%%%%%%%%%%%%%%%%%%%
\section{Conclusions and remarks} \label{sec:conclusion}
In this paper, we presented a prescription for calculating the space-time OPEs from the world-sheet in the hybrid formalism. We provided an argument to support why this proposal reproduces the space-time OPEs. We checked the OPE prescription by calculating the space-time OPEs of certain vertex operators, in particular, the generators of the $\mathcal{W}_{\infty}$ algebra in the space-time theory.

The method presented in this paper gives a direct way of reading off the space-time symmetry from the world-sheet. It explains how the world-sheet OPEs localize to the space-time OPEs. As a conclusion, we note that the (holomorphic part of the) world-sheet correlation functions that we discussed in Section~\ref{sec:correlation} calculate the space-time OPEs. Therefore, they manifestly agree with the space-time result for the untwisted sector.

Let us discuss a few possible directions for future related studies. A down-side of our analysis is that we treated a part of the world-sheet variables as fields in the path-integral formalism and the rest as honest CFT vertex operators. The analysis of \cite{Dei:2023ivl} shows that the free-field realization that we considered effectively solves the Ward identities of \cite{Eberhardt:2019ywk,Dei:2020zui}. A useful point is to understand the localization of the OPEs in the operator-formalism.

As discussed in the Introduction, we merely focused on the untwisted sector of the space-time theory. A reason for this is that there is a distinction between the $w=1$-twisted sector and $w>1$-twisted sectors. In the latter case, the space-time $w$-twisted ground states are not chiral fields, and therefore, one does not expect to naively be able to define the mode expansion that we considered. If the tensionless string is deformed by a descendant of the $2$-twisted ground states \cite{Gaberdiel:2015uca,Fiset:2022erp,Gaberdiel:2023lco}, one expects that the world-sheet is no longer at the pure NS-NS point but also has RR-flux, see \cite{Berkovits:1999im,Eberhardt:2018vho}. In particular, the huge space-time symmetry gets broken under this deformation, see \cite{Gaberdiel:2015uca}. It would be interesting to understand how this happens in our analysis on the world-sheet. A naive guess is that the fields $\bar{\mathcal{A}}(z)$ in eq.~\eqref{eq:arg-form} are no longer local fields.

In \cite{Eberhardt:2020bgq,Eberhardt:2021jvj,Aharony:2024fid} for $k=1$, it is suggested that the space-time theory is the grand canonical\- ensemble $\bigoplus_{N=0}^{\infty} e^{2\pi i \mu N} \text{Sym}^N(\mathbb{T}^4)$. In fact, in our analysis, the order of the symmetric orbi\-fold does not appear. Similarly, it was observed in \cite{Bertle:2020sgd,Gaberdiel:2021njm} that the world-sheet effectively sees a single copy of the symmetric orbifold in the untwisted sector. This suggests that the world-sheet calculates OPEs of products of operators in different copies with a fixed chemical potential $\mu$. It would be instructive to understand this point more concretely.

A particular deformation of the symmetric orbifold is the single-trace $T\overline{T}$ deformation \cite{Giveon:2017nie,Giveon:2017myj}. The analysis is usually done for $k>1$ units of NS-NS flux, and it is argued that the single trace $T\overline{T}$ connects string theory on $\text{AdS}_3$ to a $3$d asymptotically linear dilaton background. In particular, it is proposed that the deformed theory still possesses two copies of the Virasoro algebra with the same central charge as the undeformed CFT via asymptotic symmetries \cite{Georgescu:2022iyx}. One might wonder about the case $k=1$ \cite{dei:ttbar}. In this limit, which is far from the supergravity approximation, the asymptotic symmetry algebra should be formulated on the world-sheet, see \cite{Du:2024tlu}. It would be stimulating to understand whether the analysis done in this paper is useful for studying the symmetry algebra of the deformed theory.

In the previous section, we performed a sanity check related to the (anti-)commutation relations of the space-time $\mathcal{W}_{\infty}$ algebra. Our explicit calculation essentially checks that the spin-$1$ field $K$ is primary and singlet with respect to the R-symmetry currents, and that the fermions have the correct charges. However, as we mentioned, the space-time theory has a much larger symmetry called the higher spin square \cite{Gaberdiel:2015mra}. This algebra involves products of bosons and fermions and not only bilinears of them. It would be interesting to understand this symmetry from the world-sheet perspective in more detail.
%%%%%%%%%%%%%%%%
\acknowledgments
I thank Andrea Dei, Matthias Gaberdiel, Bob Knighton, Shota Komatsu, Ji Hoon Lee, Beat Nairz, Savdeep Sethi and Vit Sriprachyakul for discussions. I am especially thankful to Matthias Gaberdiel for suggesting this problem to me, insightful and important discussions, and his comments on draft versions of this article. I also thank Andrea Dei and Bob Knighton for related discussions, and their comments on a draft version of this paper. This work was supported by a grant from the Swiss National Science Foundation (SNSF) as well as the NCCR SwissMAP that is also funded by the SNSF.
\appendix
%%%%%%%%%%%%%%%%%%%%%%%%%%%%%%%%%%%%%%%%%%
\section{Some properties of the OPE prescription} \label{app:ope}
In this appendix, we discuss some properties of the OPE prescription in eq.~\eqref{eq:p-product-arg}. Let us start with showing eqs.~\eqref{eq:explicit-p-formula} and \eqref{eq:res-contour}. One expects that eq.~\eqref{eq:explicit-p-formula} holds, as it is essentially the duality relation. In order to directly show it, we use the generic form of eq.~\eqref{eq:arg-form} and write
\begin{equation} \label{eq:ope-generic-form}
	\mathcal{A}(x) = \oint dz \, \bar{\mathcal{A}}(z) \delta(\gamma(z)-x) \ ,
\end{equation}
and
\begin{equation}
	\delta(\gamma(z)-x)=\delta(\gamma(z)-y-(x-y)) = \sum_{r\in\mathbb{Z}-h_A} (\gamma(z)-y)^{r+h_A-1} (x-y)^{-r-h_A} \ ,
\end{equation}
see eq.~\eqref{eq:app-vertex-to-show} and the discussion below. Eq.~\eqref{eq:explicit-p-formula} is then equivalent to
\begin{equation} \label{eq:app-duality-to-show}
	V(\mathcal{A}_r[\gamma(z)] \ket{\mathcal{B}};y,w) = \oint_{C_w} dz \, \mathcal{A}_r[\gamma(z)-y] V(\ket{\mathcal{B}};y,w) \ ,
\end{equation}
where we are indicating the explicit dependence on $\gamma$. Using the duality relation on the world-sheet, the left-hand side equals applying $\mathcal{A}_r[\gamma(z)]$ to $V(\ket{\mathcal{B}};0,w)$ and then shifting the result to an arbitrary $y$. By inserting $e^{-y\mathcal{L}_{-1}}e^{y\mathcal{L}_{-1}}=1$ between the two operators, it equals the right-hand side of eq.~\eqref{eq:app-duality-to-show}. Another way of seeing this is as follows: as $V(\ket{\mathcal{B}};0,w)$ involves $\delta(\gamma(w))$, $\gamma(w)=0$ is imposed and therefore, $[-\gamma(w)]$ can be added to $\gamma(z)$ without changing the result. This shows that only derivatives of $\gamma(z)$ around $w$ appear, see eq.~\eqref{eq:no-x-arg}, and hence, shifting the result only shifts $\delta(\gamma(w))$ to $\delta(\gamma(w)-y)$. On the other hand, in the right-hand side of eq.~\eqref{eq:app-duality-to-show} the same idea allows us to replace $y$ with $\gamma(w)$ inside the argument of $\mathcal{A}_r$. This shows that eq.~\eqref{eq:app-duality-to-show} holds and we establish eq.~\eqref{eq:explicit-p-formula}.

Now we show eq.~\eqref{eq:res-contour}. By writing $\mathcal{S}(x,z)$ in the form given in eq.~\eqref{eq:arg-form}, and using eqs.~\eqref{eq:explicit-p-formula} and \eqref{eq:app-duality-to-show}, we see that eq.~\eqref{eq:res-contour} holds if we show
\begin{equation} \label{eq:app-res-to-show}
	\text{Res}_x \, x^{l} \delta(\gamma(z)-x) = \oint_{C_y} dx \, x^{l} \delta(\gamma(z)-y-(x-y)) \ ,
\end{equation}
for an integer $l$. This directly follows from eq.~\eqref{eq:app-vertex-res-equivalence} via setting $f(x)=x^l$ and assuming $|\gamma(z)-y|<|y|$, which is the same assumption in space-time with $|x-y|<|y|$.

As the next task, we would like to explain the statement claimed about eqs.~\eqref{eq:locality-assumption} and \eqref{eq:s-f-modes}. Let us assume that these two hold
\begin{equation} \label{eq:app-locality-assumption}
	\phi(S_r\ket{0}) = \mathcal{S}_{r} \ket{\Omega_{P=0}} \ , \quad r\in\mathbb{Z}-h_S \ ,
\end{equation}
\begin{equation} \label{eq:app-s-f-modes}
	\bar{\phi}([S_r,F_s])=[\mathcal{S}_r,\mathcal{F}_s] \ ,
\end{equation}
and recursively argue that eq.~\eqref{eq:arg-mode-to-show} is satisfied
\begin{equation} \label{eq:app-mode-to-show}
	\phi(S_r \ket{T}) = \mathcal{S}_r \ket{\mathcal{T}} \ , \quad r\in \mathbb{Z}-h_S \ .
\end{equation}
If $\ket{T}=\ket{0}$, using the definition of the map $\phi$ and eq.~\eqref{eq:app-locality-assumption}, we see that eq.~\eqref{eq:app-mode-to-show} is satisfied for any physical field $\mathcal{S}$. Now we assume that it holds for any $\ket{T}$ that involves $k\geq 0$ free fields, see eq.~\eqref{eq:st-generic-state}. We want to argue that eq.~\eqref{eq:app-mode-to-show} is satisfied for a state that has $k+1$ free fields by writing the left-hand side of this equation for
\begin{equation} \label{eq:app-mid-1}
	\phi(S_r F_{-s} \ket{T}) = \phi([S_r,F_{-s}]\ket{T}+(-1)^{f_F f_S} F_{-s} S_{r}\ket{T}) \ , \quad (s>0) \ .
\end{equation}
We set
\begin{equation} \label{eq:app-arg-3}
	[S_r,F_l] = \sum_{n=0}^M Q^n_{r+l} \ , \quad [\mathcal{S}_r,\mathcal{F}_l] = \sum_{n=0}^M \mathcal{Q}^n_{r+l} \ ,
\end{equation}
where we have assumed eq.~\eqref{eq:app-s-f-modes} and used eq.~\eqref{eq:def-phi} for $\bar{\phi}$. By assumption, eq.~\eqref{eq:app-mode-to-show} holds for $\ket{T}$ and any arbitrary $S$. Therefore, for the first term on the right-hand side of eq.~\eqref{eq:app-mid-1} we get
\begin{equation}
	\phi([S_r,F_{-s}]\ket{T})=[\mathcal{S}_r,\mathcal{F}_{-s}] \ket{\mathcal{T}} \ .
\end{equation}
For the second term, we write
\begin{equation}
	S_r \ket{T} = \sum_{n=0}^N \ket{O^n} \ , \quad \mathcal{S}_r \ket{\mathcal{T}} = \sum_{n=0}^N \ket{\mathcal{O}^n} \ ,
\end{equation}
where we have used that eq.~\eqref{eq:app-mode-to-show} holds for $\ket{T}$. Therefore, we have
\begin{equation}
	\phi(F_{-s} S_r \ket{T}) = \sum_{n=0}^N \phi(F_{-s}\ket{O^n}) = \sum_{n=0}^N \mathcal{F}_{-s} \ket{\mathcal{O}^n} =\mathcal{F}_{-s} \mathcal{S}_r \ket{\mathcal{T}} \ ,
\end{equation}
where $s>0$. Note that in the second equality, we used the definition of the map $\phi$, see eq.~\eqref{eq:def-phi-generic-state}. This shows that eq.~\eqref{eq:app-mode-to-show} is satisfied once eqs.~\eqref{eq:app-locality-assumption} and \eqref{eq:app-s-f-modes} hold. Let us now discuss that eq.~\eqref{eq:app-locality-assumption} holds for all $r\in\mathbb{Z}-h_S$ if it holds for $r\geq -h_S$. Indeed, we argue recursively that eq.~\eqref{eq:app-locality-assumption} holds for $r=-h_S-l$ for $l\geq 0$. For $l=0$, this is an assumption. We assume that it holds for $l\geq 0$ and show
\begin{equation} \label{eq:app-l-ind}
	\phi(S_{-h_S-l-1} \ket{0}) = \mathcal{S}_{-h_S-l-1} \ket{\Omega_{P=0}} \ .
\end{equation}
We start by writing
\begin{equation}
	\phi(S_{-h_S-l-1} \ket{0}) = \frac{1}{l+1}\phi(L_{-1} S_{-h_S-l} \ket{0}) \ ,
\end{equation}
where $L_{-1}$ here indicates the space-time translation operator. This relation also holds on the world-sheet because
\begin{equation} \label{eq:derivative-state}
	\partial_x V(\ket{\mathcal{A}};x,z) = V(\mathcal{L}_{-1} \ket{\mathcal{A}};x,z) \ ,
\end{equation}
where $\mathcal{L}_{-1}=J^+_0$. This follows from the definition \eqref{eq:vertex-operator-x-z} and the duality relation on the world-sheet. Since $\mathcal{L}_{-1} \ket{\mathcal{A}}$ will be of the general form in eq.~\eqref{eq:generic-state-arg}, see eqs.~\eqref{eq:app-lm1-comms} and \eqref{eq:app-lm1-brst} below, the argument for eq.~\eqref{eq:locality-assumption} still holds with $r=-h_{A}-1$ and we recursively get
\begin{equation}
	\mathcal{A}_{-h_A-l} \ket{\Omega_{P=0}} = \frac{1}{l!} (\mathcal{L}_{-1})^l \mathcal{A}_{-h_A} \ket{\Omega_{P=0}} \ ,
\end{equation}
up to BRST exact terms. This gives
\begin{equation} \label{eq:app-sl-derivative}
	\mathcal{S}_{-h_S-l-1} \ket{\Omega_{P=0}} = \frac{1}{l+1} \mathcal{L}_{-1} \mathcal{S}_{-h_S-l} \ket{\Omega_{P=0}} \ .
\end{equation}
Now we write
\begin{equation}
	S_{-h_S-l} \ket{0} = \sum_{n=0}^N \ket{O^n} \ , \quad \mathcal{S}_{-h_S-l} \ket{\Omega_{P=0}} = \sum_{n=0}^N \ket{\mathcal{O}^n} \ ,
\end{equation}
where we used that eq.~\eqref{eq:app-locality-assumption} holds for $r=-h_S-l$. Therefore, it is enough to show
\begin{equation} \label{eq:l1-app}
	\phi(L_{-1} \ket{O}) = \mathcal{L}_{-1} \ket{\mathcal{O}} \ .
\end{equation}
In order to show this, we write
\begin{equation}
	\ket{O}=F^{(1)}_{-s_1} \cdots F^{(m)}_{-s_m} \ket{0} \ , \quad \ket{\mathcal{O}} =\mathcal{F}^{(1)}_{-s_1} \cdots \mathcal{F}^{(m)}_{-s_m} \ket{\Omega_{P=0}} \ .
\end{equation}
Now we note that
\begin{equation} \label{eq:app-lm1-comms}
	[L_{-1},F^{(j)}_{p_j}] = (1-h_j-p_j) F^{(j)}_{p_j-1} \ , \quad [\mathcal{L}_{-1},\mathcal{F}^{(j)}_{p_j}] = (1-h_j-p_j) \mathcal{F}^{(j)}_{p_j-1} \ ,
\end{equation}
where the second relation is shown in \cite{Naderi:2022bus}. Moreover, we have \cite{Naderi:2022bus}
\begin{equation} \label{eq:app-lm1-brst}
	L_{-1} \ket{0} = 0 \ , \quad \mathcal{L}_{-1} \ket{\Omega_{P=0}} = \text{BRST exact} \ .
\end{equation}
This subsequently shows that eqs.~\eqref{eq:l1-app} and \eqref{eq:app-l-ind} hold. We conclude that it is enough to check eq.~\eqref{eq:app-locality-assumption} for $r\geq -h_S$. In particular, note that if eq.~\eqref{eq:app-locality-assumption} holds for $r\geq -h_S$, then
\begin{equation} \label{eq:def-vertex-st}
	\phi(e^{x L_{-1}} S_{-h_S}\ket{0}) = e^{x\mathcal{L}_{-1}} \mathcal{S}_{-h_S} \ket{\Omega_{P=0}} \ .
\end{equation}
We want to study how the map $P$ in eq.~\eqref{eq:p-product-arg} behaves under exchanging $\mathcal{A}$ and $\mathcal{B}$. As we mentioned, $\delta(\gamma(z)-x)$ is a non-local field on the world-sheet. For this reason, we find it easier to study this question indirectly using the results of Section~\ref{sec:argument}. In that section, we showed that the OPE calculations using the map $P$ coincide with the space-time OPE calculations. As the space-time OPE is local, this shows
\begin{equation} \label{eq:app-exchange}
	P[\mathcal{A},\mathcal{B}](x,y;w) = (-1)^{f_A f_B} P[\mathcal{B},\mathcal{A}](y,x;w) \ ,
\end{equation}
up to BRST exact terms. Note that this relation should be read as one side being the analytic continuation of the other side, see \cite{Gaberdiel:1998fs,Goddard:1989dp,Kac:1996wd}. The parity $f_X$ that we discuss in eq.~\eqref{eq:arg-ax} is defined via treating $\gamma$ as a bosonic field. It is then straightforward to show that the space-time modes of $\mathcal{A}(x,z)$ associated with a space-time vertex operator $A(x)$ are bosonic (fermionic) on the world-sheet if $A$ is bosonic (fermionic) in the space-time, see eq.~\eqref{eq:generic-state-arg} and eq.~\eqref{eq:ground-state-pictures}. Note that however, we are suppressing the `cocycle factors' in front of $\delta(\gamma(z)-x)$ in the left-hand side of the identity \eqref{eq:distribution-equality} that are needed for locality \cite{Goddard:1986ts}.

Let us explain that in the definition given in eq.~\eqref{eq:p-product-arg}, it does not matter how one distributes the picture number between the two fields. This follows because the operators in eq.~\eqref{eq:arg-ax} are DDF operators. In fact, it is known that it does not matter how one distributes the picture number between DDF operators and the state that they act on, see e.g.\ Appendix~A of \cite{Naderi:2022bus}. Let us explain this briefly. Assume that there is a DDF operator $D$ written as follows
\begin{equation}
	D = \oint dz \, F(z) = F_0 \ ,
\end{equation}
where $F$ is primary with weight $1$ on the world-sheet. We define the picture-raising operator on $D$ as
\begin{equation}
	P_+ D = [G^+_0,(\xi F)_0] \ ,
\end{equation}
whereby $(\xi F)_0$ we mean the radial normal-ordering, see eq.~\eqref{eq:picture-raising}. Then up to BRST exact states, we have
\begin{equation} \label{eq:p-raising-app}
	(P_+ D) \psi = D (P_+ \psi) = P_+ (D \psi) \ ,
\end{equation}
where $\psi$ is a physical state, see \cite{Naderi:2022bus}. By writing $D=\mathcal{A}(x-y)$ and $\psi=\ket{\mathcal{B}}$ in eq.~\eqref{eq:p-raising-app}, then we see that it does not matter how one distributes the picture number in eq.~\eqref{eq:p-product-arg}.

Lastly, we note that if we have calculated the OPE $P[\mathcal{A},\mathcal{B}](x,y;w)$ in eq.~\eqref{eq:p-product-arg}, then the OPE $P[\partial_x \mathcal{A},\mathcal{B}](x,y;w)$ is obtained simply by taking a derivative with respect to $x$ from the expression in $P[\mathcal{A},\mathcal{B}](x,y;w)$, as one expects from the space-time point of view. One way of seeing this is that the space-time mode expansion of the derivative field $\partial_x \mathcal{A}(x,z)$ is obtained by simply taking an $x$-derivative from the mode expansion of $\mathcal{A}(x,z)$, see eq.~\eqref{eq:derivative-state} and the discussion below. Therefore, $P[\partial_x \mathcal{A},\mathcal{B}](x,y;w)$ is obtained from $P[\mathcal{A},\mathcal{B}](x,y;w)$ via taking an $x$-derivative of $P[\mathcal{A},\mathcal{B}](x,y;w)$.
%%%%%%%%%%%%%%%%%%%%%%%%%%%%%%%%%%%%%%%%%%%%%%%%%%%%%%%%%%
\section{Formal distributions} \label{app:vertex}
In this appendix, we briefly review formal distributions following \cite{Kac:1996wd}. Let $V$ be a $\mathbb{Z}_2$-graded vector space such that $V=V_0\oplus V_1$ and set $U=\text{End}(V)$. A formal distribution with values in $U$ is defined as a formal series in finitely many variables
\begin{equation}
	a(z_1,\cdots,z_k) = \sum_{n_j\in \mathbb{Z}} a_{n_1,\cdots,n_k} z_1^{n_1} \cdots z_k^{n_k} \ , \quad (k\geq 1) \ .
\end{equation}
We define the parity of $a$ as follows: if for all $n_j \in \mathbb{Z}$, $a_{n_1,\cdots,n_k}$ maps $V_{\alpha}$ to $V_{\alpha+p(a)}$ for a constant $p(a)$, $p(a)$ is the parity of $a$ in $\mathbb{Z}_2$. We say that $a$ is bosonic (fermionic) if $p(a)=0 \bmod 2$ ($p(a)=1 \bmod 2$). The derivative of $a$ is defined naturally as
\begin{equation}
	\partial_{z_l} a(z_1,\cdots,z_k) = \sum_{n_j\in \mathbb{Z}} n_l z_l^{-1} a_{n_1,\cdots,n_k} z_1^{n_1} \cdots z_k^{n_k} \ .
\end{equation}
Let us now focus on a formal distribution in a single variable
\begin{equation}
	a(z) = \sum_{n\in\mathbb{Z}} a_n z^n \ .
\end{equation}
In this case, one defines the `residue' as the coefficient of $z^{-1}$, i.e.\ $\text{Res}_z a(z)=a_{-1}$. Since $\text{Res}_z \partial_z a(z)=0$, we have
\begin{equation}
	\text{Res}_z [\partial_z a(z) b(z)] = - \text{Res}_z [a(z) \partial_z b(z)] \ ,
\end{equation}
provided that the product of $a(z)$ and $b(z)$ is defined, e.g.\ one of them is a Laurent polynomial. Moreover, we can consider the Taylor expansion of formal distributions. In fact, one can show that
\begin{equation} \label{eq:taylor-formula}
	a(z) = \sum_{n=0}^{\infty} \frac{\partial^n_w a(w)}{n!} (z-w)^n \ ,
\end{equation}
where $|z-w|<|w|$, see Proposition~2.4 in \cite{Kac:1996wd} for more details. We define the formal delta-function as follows
\begin{equation} \label{eq:app-def-delta-function}
	\delta(z-w) = \sum_{n\in\mathbb{Z}} w^{n} z^{-n-1} \ .
\end{equation}
Assume that $f$ is a distribution. Then the product of $f$ with the formal delta-function is defined and $\text{Res}_z f(z) \delta(z-w) = f(w)$. Therefore, it behaves as a normal delta-function. Moreover, one can show that $f(z)\delta(z-w)=f(w)\delta(z-w)$. A natural question is how this definition is related to the Taylor expansion of $1/(z-w)$ when e.g.\ $|z|>|w|$? Following \cite{Kac:1996wd}, let us denote 
\begin{equation}
	i_{|z|>|w|} \frac{1}{z-w} = \sum_{n=0}^{\infty} w^{n} z^{-n-1} \ , \quad (|z|>|w|) \ ,
\end{equation}
which is the Taylor expansion in the region $|z|>|w|$. Comparing the coefficients, it is straightforward to show that
\begin{equation} \label{eq:delta-one-over}
	\delta(z-w) = i_{|z|>|w|} \frac{1}{z-w} + i_{|w|>|z|} \frac{1}{w-z} \ .
\end{equation}
This means that the formal delta-function should be seen as a sum of Taylor expansions of $1/(z-w)$ in different regions. In fact, using eq.~\eqref{eq:delta-one-over}, it is easy to show
\begin{equation}
	\text{Res}_x \, f(x) \delta(z-x) = \oint_{C_0} dx \, f(x) \delta(z-x) \ ,
\end{equation}
where $C_p$ is a contour around $p$. We list a few properties of the formal delta-function:
\begin{itemize}
	\item $\delta(z-w) = \delta(w-z)$,
	\item $\partial_z^n \delta(z-w) = (-1)^n \partial_w^n \delta(z-w) \ , \quad (n\geq 0)$,
	\item $(z-w)^{n+1} \partial_z^n \delta(z-w) = 0 \ , \quad (n\geq 0)$.
\end{itemize}
These properties can be directly shown using the definition of the formal delta-function. In what sense the delta-function $\delta(z-w)$ only depends on $(z-w)$? Given an arbitrary $a$, let us consider the delta-function $\delta(z_a-x_a)$ with $y_a:=y-a$. What we mean by this is substituting $z_a$ and $x_a$ as formal variables in eq.~\eqref{eq:app-def-delta-function} and consider
\begin{equation} \label{eq:app-vertex-to-show}
	\text{Res}_x \, f(x) \delta(z-x) = \begin{cases}
		\text{Res}_t \, f(a+t) \delta(z_a-t) \ , & \quad |z-a|<|a|\\
		\text{Res}_t \, f(t+a) \delta(z_a-t) \ , & \quad |z-a|>|a| 
	\end{cases} \ .
\end{equation}
Let us show this equality for instance when $|z-a|<|a|$. In this case, using the Taylor expansion for $f$, see eq.~\eqref{eq:taylor-formula}, we have
\begin{equation}
	f(a+t) = \sum_{l=0}^{\infty} \frac{\partial^l f(a)}{l!} t^l \ ,
\end{equation}
for $|t|<|a|$. Taking the residue in the right-hand side of eq.~\eqref{eq:app-vertex-to-show}, we get
\begin{equation}
	\text{Res}_t \, f(a+t) \delta(z_a-t) = \sum_{l=0}^{\infty} \frac{\partial^l f(a)}{l!} (z-a)^l =f(z) \ ,
\end{equation}
for $|z-a|<|a|$. A similar argument applies for the case $|z-a|>|a|$. Note that eq.~\eqref{eq:app-vertex-to-show} can be more compactly written as
\begin{equation} \label{eq:app-vertex-res-equivalence}
	\text{Res}_x \, f(x) \delta(z-x) = \oint_{C_a} dx \, f(x) \delta(z_a-x_a) \ .
\end{equation}
Now let us consider formal distributions in $2$ variables $z$ and $w$
\begin{equation}
	a(z,w) = \sum_{n,m\in\mathbb{Z}} a_{n,m} z^n w^m \ .
\end{equation}
We define
\begin{equation}
	a(z,w)^{(+),z}=\sum_{n\in\mathbb{Z}_{\geq 0},m\in\mathbb{Z}} a_{n,m} z^n w^m \ .
\end{equation}
It can be shown that any formal distribution $a(z,w)$ can be uniquely represented as
\begin{equation}
	a(z,w) = \sum_{n=0}^{\infty} \frac{c^n(w)}{n!} \partial_w^n \delta(z-w) + b(z,w) \ , \quad c^n(w) = \text{Res}_z a(z,w) (z-w)^n \ ,
\end{equation}
where $b(z,w)$ does not include negative powers of $z$, i.e.\ $b(z,w)=b(z,w)^{(+),z}$. Having this, we call two distributions $a$ and $b$ mutually local if there exists an $N\in \mathbb{Z}_{\geq 0}$ such that
\begin{equation} \label{eq:app-locality}
	(z-w)^{N+1} [a(z),b(w)] = 0 \ ,
\end{equation}
where the bracket is the super-commutator, i.e.\ commutator if at least one of the operators is bosonic and anti-commutator if both are fermionic. Before we state the main result, let us set
\begin{equation}
	a(z)_+ = \sum_{n\leq -1} a_n z^{-n-1} \ , \quad a(z)_- = \sum_{n>-1} a_n z^{-n-1} \ ,
\end{equation}
and define
\begin{equation} \label{eq:app-normal}
	:a(z)b(w): = a(z)_+ b(w) + (-1)^{p(a)p(b)} b(w) a(z)_- \ .
\end{equation}
The condition that two distributions are mutually local with a given $N$ in eq.~\eqref{eq:app-locality} is equivalent to
\begin{equation} \label{eq:app-super}
	[a(z),b(w)] = \sum_{n=0}^{N} \frac{1}{n!} \partial_w^n \delta(z-w) c^n(w) \ ,
\end{equation}
where $c^j(w)$ are formal distributions. It means there are finitely many singular terms in the (anti-)commutator. This is equivalent to
\begin{equation} \label{eq:app-ope-vertex}
	a(z)b(w) = \sum_{n=0}^{N} i_{|z|>|w|} \frac{1}{(z-w)^{n+1}} c^n(w) + :a(z)b(w): \ ,
\end{equation}
which is the usual operator product expansion (OPE). What is more, all these conditions are also equivalent to
\begin{equation} \label{eq:app-statistic}
	(-1)^{p(a)p(b)}b(w)a(z) = \sum_{n=0}^{N} i_{|w|>|z|} \frac{1}{(z-w)^{n+1}} c^n(w) + :a(z)b(w): \ ,
\end{equation}
\begin{equation} \label{eq:app-comm-finite}
	[a_n,b_m]=\sum_{j=0}^{N} {n \choose j} c^j_{n+m-j} \ .
\end{equation}
The last condition means that there are finitely many terms in the (anti-)commutator of the modes, see Theorem~2.3 in \cite{Kac:1996wd} for more details.
%%%%%%%%%%%%%%%%%%%%%%%%%%%%%%%%%%%%%%%%%%%%%%%%%%%%%
\section{The superconformal algebras} \label{app:sca}
In this appendix, we fix our conventions for the superconformal algebras and their topolo\-gically twisted versions. In particular, we consider a small $\mathcal{N}=4$ superconformal algebra with central charge $c=6$. The fields are the stress-tensor $T$, the R-symmetry currents $J$, $J^{++}$ and $J^{--}$, and the supercurrents $G^+$, $G^-$, $\tilde{G}^+$ and $\tilde{G}^-$. In general, the number of `$+$' and `$-$' specifies twice the charge under $J$, for example, $G^+$ has charge $(+1/2)$ while $J^{--}$ has charge $(-1)$. The R-symmetry currents form an $\mathfrak{su}(2)_1$ and they are primary fields of weight $1$
\begin{subequations} \label{eq:sca-r-symmetry}
	\begin{equation}
		J^+(z) J^-(w) \sim \frac{1}{(z-w)^2}+\frac{2 J(w)}{(z-w)} \ ,
	\end{equation}
	\begin{equation}
		J(z) J(w) \sim \frac{c/12}{(z-w)^2} \ .
	\end{equation}
\end{subequations}
The supercurrents are all primary with weight $(3/2)$. They satisfy
\begin{subequations} \label{eq:sca-supercurrents}
	\begin{equation}
		G^+(z) G^-(w) \sim \frac{c/3}{(z-w)^3}+\frac{2 J(w)}{(z-w)^2}+\frac{T(w)+\partial J(w)}{(z-w)} \ ,
	\end{equation}
	\begin{equation}
		\tilde{G}^+(z) \tilde{G}^-(w) \sim \frac{c/3}{(z-w)^3}+\frac{2J(w)}{(z-w)^2}+\frac{T(w)+\partial J(w)}{(z-w)} \ ,
	\end{equation}
	\begin{equation}
		J^{\pm\pm}(z) G^{\mp}(w) \sim \frac{\pm \tilde{G}^{\pm}(w)}{(z-w)} \ ,
	\end{equation}
	\begin{equation}
		J^{\pm\pm}(z) \tilde{G}^{\mp}(w) \sim \frac{\mp G^{\pm}(w)}{(z-w)} \ ,
	\end{equation}
	\begin{equation}
		G^{\pm}(z) \tilde{G}^{\pm}(w) \sim \frac{\mp 2 J^{\pm\pm}(w)}{(z-w)^2}+\frac{\mp \partial J^{\pm\pm}(w)}{(z-w)} \ .
	\end{equation}
\end{subequations}
The topological twist amounts to shifting the stress-tensor by $T \rightarrow T\pm \partial J$. Let us focus on the positive choice here as this is relevant for us, see Appendix~\ref{app:hybrid}. We will denote the fields of the topologically twisted theory by a subscript `C'. With this choice, for a field $F_C$ that had weight $\Delta_0$ and charge $q$ before the topological twist, now it will have weight $\Delta_{\text{twist}}=\Delta_0-q$. For instance, $G^+_C$ has weight $1$ while $J^{--}_C$ has weight $2$. All the OPEs remain unchanged except
\begin{subequations} \label{eq:sca-top-twisted}
	\begin{equation}
		G^+_C(z) G^-_C(w) \sim \frac{c/3}{(z-w)^3}+\frac{2 J_C(w)}{(z-w)^2}+\frac{T_C(w)}{(z-w)} \ ,
	\end{equation}
	\begin{equation}
		\tilde{G}^+_C(z) \tilde{G}^-_C(w) \sim \frac{c/3}{(z-w)^3}+\frac{2J_C(w)}{(z-w)^2}+\frac{T_C(w)}{(z-w)} \ ,
	\end{equation}
	\begin{equation}
		T_C(z) J_C(w) \sim \frac{-c/6}{(z-w)^3}+\frac{J_C(w)}{(z-w)^2}+\frac{\partial J_C(w)}{(z-w)} \ .
	\end{equation}
\end{subequations}
%%%%%%%%%%%%%%%%%%%%%%%%%%%%%%%%%%%%%%%%%%%%%%%%%
\section{The \texorpdfstring{$\mathbb{T}^4$}{T4} theory} \label{app:t4}
In this appendix, we fix our conventions for the $\mathbb{T}^4$ CFT. This theory appears both in space-time and on the world-sheet. Let us first describe the $\mathbb{T}^4$ of the space-time. It is convenient to describe it in terms of complex free bosons and free fermions
\begin{equation} \label{eq:bos-ferm-t4-ope}
	\partial X^j(z) \partial \bar{X}^k(w) \sim \frac{\delta^{jk}}{(z-w)^2} \ , \quad \psi^{\alpha,j}(z) \psi^{\beta,k}(w) \sim \frac{\epsilon^{\alpha\beta}\epsilon^{kj}}{(z-w)} \ ,
\end{equation}
where $j,k\in\{1,2\}$, $\alpha,\beta\in\{+,-\}$, $\epsilon^{+-}=-\epsilon^{-+}=1$, $\epsilon^{12}=-\epsilon^{21}=1$, and the other combinations vanish. The $4$ free bosons and $4$ free fermions form an $\mathcal{N}=2$ superconformal algebra with $c=6$ where
\begin{subequations} \label{eq:t4-n=2}
	\begin{equation}
		T = \partial X^j \partial \bar{X}^j + \frac{1}{2} \epsilon^{\alpha\beta} \epsilon^{jk} \psi^{\alpha,j} \partial \psi^{\beta,k} \ ,
	\end{equation}
	\begin{equation}
		J = -\frac{1}{4} \delta^{\alpha,-\beta} \epsilon^{jk} \psi^{\alpha,j} \psi^{\beta,k} \ ,
	\end{equation}
	\begin{equation} \label{eq:app-t4-gp}
		G^+ = \partial\bar{X}^j \psi^{+,j} \ ,
	\end{equation}
	\begin{equation}
		G^- = - \epsilon^{jk} \partial X^j \psi^{-,k} \ ,
	\end{equation}
\end{subequations}
see Appendix~\ref{app:sca}. The $4$ free fermions realize $\mathfrak{su}(2)_1 \oplus \mathfrak{su}(2)_1$. Considering only one of the $\mathfrak{su}(2)_1$, it is realized by $J$, $J^{++}$ and $J^{--}$ where
\begin{equation} \label{eq:r-symmetry-jpp-jmm}
	J^{++} = \psi^{+,1} \psi^{+,2} \ , \quad J^{--}=\psi^{-,2} \psi^{-,1} \ .
\end{equation}
Together with the generators in eqs.~\eqref{eq:t4-n=2}, they form a small $\mathcal{N}=4$ superconformal algebra with $c=6$ where the additional supercurrents are
\begin{subequations}
	\begin{equation}
		\tilde{G}^+ = - \epsilon^{jk} \partial X^j \psi^{+,k} \ ,
	\end{equation}
	\begin{equation}
		\tilde{G}^- = - \partial \bar{X}^j \psi^{-,j} \ .
	\end{equation}
\end{subequations}
The $\mathbb{T}^4$ also appears on the world-sheet as a topologically twisted theory, see Appendix~\ref{app:sca}. We have denoted the generators of this topologically twisted $\mathcal{N}=4$ algebra with a subscript `C' on the world-sheet. It is useful to bosonize the fermions on the world-sheet as follows
\begin{equation}
	\psi^{+,j} = e^{iH^j} \ , \quad \psi^{-,j}=\epsilon^{jk} e^{-iH^k} \ ,
\end{equation}
where $H^j$ are bosons satisfying
\begin{equation}
	H^j(z) H^k(w) \sim -\delta^{jk} \ln{(z-w)} \ ,
\end{equation}
and the stress-tensor is
\begin{equation}
	T_{H_1,H_2} = -\frac{1}{2} [(\partial H^1)^2+(\partial H^2)^2]+\frac{1}{2} \partial^2(iH^1+iH^2) \ .
\end{equation}
We will also denote
\begin{equation} \label{eq:def-h}
	H=H^1+H^2 \ .
\end{equation}
%%%%%%%%%%%%%%%%%%%%%%%%%%%%%%%%%%%%%%%%%%%%%%%%%
\section{The hybrid formalism} \label{app:hybrid}
In this appendix, we briefly review the hybrid formalism of BVW \cite{Berkovits:1999im}. There are a few recent reviews on this topic, see e.g.\ \cite{Gerigk:2012lqa,Naderi:2022bus,Demulder:2023bux}. The fields that appear on the world-sheet are reviewed in Section~\ref{sec:review}. The world-sheet theory is described by the following theories
\begin{equation} \label{eq:hybrid-fields}
	\mathfrak{psu}(1,1|2)_1 \oplus (\rho,\sigma) \oplus (\text{a topologically twisted } \mathbb{T}^4) \ .
\end{equation}
The $\mathfrak{psu}(1,1|2)_1$ is discussed in Section~\ref{sec:psu}, and the topologically twisted $\mathbb{T}^4$ is reviewed in Appendix~\ref{app:t4}. The $\rho$ and $\sigma$ are bosons that satisfy
\begin{equation}
	A(z) A(w) \sim -\ln{(z-w)} \ , \quad A\in\{\rho,\sigma\} \ ,
\end{equation}
and have a trivial OPE between each other. Their stress-tensor is
\begin{equation}
	T_{\rho\sigma}=-\frac{1}{2} [(\partial\rho)^2+(\partial\sigma)^2] + \frac{3}{2} \partial^2(\rho+i\sigma) \ .
\end{equation}
As we mentioned in Section~\ref{sec:review}, the ingredients in \eqref{eq:hybrid-fields} combine to form a topologically twisted $\mathcal{N}=4$ algebra with $c=6$ on the world-sheet, see Appendix~\ref{app:sca}. The $\mathcal{N}=2$ generators are
\begin{subequations} \label{eq:hybrid-n=2}
	\begin{equation} \label{eq:hybrid-stress-tensor}
		T=T_{\text{free}}+T_{\rho\sigma}+T_C \ ,
	\end{equation}
	\begin{equation}
		G^+ = e^{-\rho} Q + e^{i\sigma} T - \partial[e^{i\sigma} \partial(\rho+iH)]+G^+_C \ ,
	\end{equation}
	\begin{equation}
		G^-=e^{-i\sigma} \ ,
	\end{equation}
	\begin{equation}
		J = \frac{1}{2} \partial(\rho+i\sigma+iH) \ ,
	\end{equation}
\end{subequations}
where
\begin{equation} \label{eq:psu-stress-tensor}
	T_{\text{free}} = -(\beta \partial \gamma)-(p_j \partial \theta^j) \ ,
\end{equation}
\begin{equation} \label{eq:hybrid-q}
	Q = p_1 p_2 \partial \gamma \ .
\end{equation}
The fields that have a subscript `C' are the fields of the topologically twisted $\mathbb{T}^4$ on the world-sheet, see Appendix~\ref{app:t4}. $H$ is defined in eq.~\eqref{eq:def-h}. The algebra in eqs.~\eqref{eq:hybrid-n=2} can be enhanced to a small $\mathcal{N}=4$ superconformal algebra: the R-symmetry currents are
\begin{equation} \label{eq:hybrid-r-symmetry}
	J^{++} = e^{\rho+i\sigma+iH} \ , \quad J^{--} = e^{-\rho-i\sigma-iH} \ .
\end{equation}
The additional supercurrents take the form
\begin{subequations}
	\begin{equation}
		\tilde{G}^+ = e^{\rho+iH} \ ,
	\end{equation}
	\begin{equation}
		\tilde{G}^- = e^{-2\rho-i\sigma-iH} Q - e^{-\rho-iH} T - e^{-\rho-i\sigma} \tilde{G}^-_C + e^{-\rho-iH} [\partial(i\sigma)\partial(\rho+iH)+\partial^2(\rho+iH)] \ .
	\end{equation}
\end{subequations}
The physical state conditions are defined in eq.~\eqref{eq:physical}. However, given a physical state, there is a hierarchy of equivalent representations of the same physical state, which we refer to as being in different `pictures'. Let us assume $\psi$ is a physical state, i.e.\ it satisfies eq.~\eqref{eq:physical}. The picture of $\psi$, which we generally denote as $P$ when the state $\psi$ is clearly understood, is defined as the eigenvalue of $\psi$ under the current $P_{\rho}=\partial \rho$. We define the picture-raising operator as \cite{Friedan:1985ey,Friedan:1986rx}
\begin{equation} \label{eq:picture-raising}
	P_+ = G^+_0 \xi_0 \ , \quad \xi = -e^{-\rho-iH} \ .
\end{equation}
It can be shown that $P_+ \psi$ is physical and its picture number is increased. Also, it does not matter how one distributes picture numbers between different states in an $n$-point function, as long as the sum of the picture numbers is fixed to be
\begin{equation}
	\sum_{j=1}^n P_j = -4 \ ,
\end{equation}
on the world-sheet sphere, see \cite{Friedan:1985ey,Berkovits:1994vy,Lust:1989tj,Gaberdiel:2021njm} and the discussion in Section~\ref{sec:correlation}.
%%%%%%%%%%%%%%%%%%%%%%%%%%%%%%%%%%%%%%%%%%%%%%%%%%%
\section{Examples of OPE calculations} \label{app:example}
In this appendix, we discuss two non-trivial examples of calculating space-time OPEs from the world-sheet, together with an example of how our argument in Section~\ref{sec:argument} works. In particular, we first consider two supercurrents and calculate their OPE, and then use this to calculate the OPE of the stress-tensor with itself.
%%%%%%%%%%%%%%%%%%%%%%%%%%
\subsection{Supercurrents}
The vertex operators that correspond to half of the supercurrents $G^+$ and $G^-$ are \cite{Gaberdiel:2021njm}
\begin{subequations} \label{eq:app-supercurrents-fields}
\begin{equation} \label{eq:app-examples-gp}
	\mathcal{G}^+(x,z) = -e^{if_2} \delta(\gamma(z)-x) e^{-\rho+i\sigma} \tilde{G}^-_C \ ,
\end{equation}
\begin{equation}
	\mathcal{G}^-(x,z) = e^{2if_1-3if_2} (\partial\gamma)^{-3} \delta(\gamma(z)-x) e^{5\rho+i\sigma+3iH} G^-_C \ ,
\end{equation}	
\end{subequations}
see Appendix~\ref{app:t4}. Let us consider the anti-commutator of the DDF operators associated with these two supercurrents. This was calculated in \cite{Naderi:2022bus}
\begin{equation} \label{eq:twisted-stress-tensor}
	\{\mathcal{G}^{+,P=1}_{-1/2},\mathcal{G}^-_{-3/2}\} \Omega_{P=-6}(x,z) = e^{2if_1-2if_2} (\partial \gamma)^{-3} \delta(\gamma(z)-x) e^{4\rho+i\sigma+2iH} D \ ,
\end{equation}
and we find
\begin{equation} \label{eq:d-stress-tensor}
	D = \sum_{j=1}^2 \partial X^j \partial \bar{X}^j + \frac{1}{2} (\partial l_j)^2 + \frac{1}{2} \partial^2 l_j \ , \quad l_j = if_2-\rho-iH^j \ .
\end{equation}
From the space-time point of view, this anti-commutator is equal to the topologically twisted stress-tensor $T_C=T+\partial J$, see Appendix~\ref{app:t4}. In \cite{Naderi:2022bus}, it is shown that the state associated with the stress-tensor and the one obtained from eq.~\eqref{eq:twisted-stress-tensor} after `undoing' the topological twist, agree up to BRST exact states. Having this, we find it useful to work with the topologically twisted stress-tensor in eq.~\eqref{eq:twisted-stress-tensor}, and we denote the associated world-sheet vertex operator as
\begin{equation} \label{eq:topologically-twisted-state}
	\mathcal{T}_{P=-4}(x,z)=e^{2if_1-2if_2} (\partial \gamma)^{-3} \delta(\gamma(z)-x) e^{4\rho+i\sigma+2iH} D \ .
\end{equation}
We would like to calculate the OPE of two supercurrents using eq.~\eqref{eq:p-product-arg}. This OPE calculation is in essence similar to the calculations discussed in Section~\ref{sec:checks} but rather more complicated. We substitute the supercurrents in eqs.~\eqref{eq:app-supercurrents-fields} into eq.~\eqref{eq:p-product-arg} and calculate the world-sheet OPE. We write the delta-function $\delta(\gamma(z)-x)$ as $1/(x-\gamma(z))$ for $|x|>|\gamma(z)|$, see eq.~\eqref{eq:delta-one-over} and the beginning of Section~\ref{sec:checks}, and take advantage of the appearance of the second delta-function and the chain rules, similar to eqs.~\eqref{eq:delta-impose} and \eqref{eq:chain-rule}. After a tedious calculation using eq.~\eqref{eq:explicit-p-formula}, we get
\begin{equation}
	P[\mathcal{G}^+,\mathcal{G}^-](x,y;w) \sim \frac{\mathcal{T}_{P=-4}(y,w)}{x-y} + \frac{\mathcal{A}(y,w)}{(x-y)^2}+\mathcal{B}(y,w) \partial_w^2 \delta(\gamma(w)-x) \ ,
\end{equation}
where we have defined
\begin{equation}
	\mathcal{A}(y,w) = e^{2if_1-2if_2} (\partial \gamma)^{-2} \delta(\gamma(w)-y) e^{4\rho+i\sigma+2iH} \partial[2if_2-2\rho-iH] \ ,
\end{equation}
\begin{equation}
	\mathcal{B}(y,w) = e^{2if_1-2if_2} (\partial \gamma)^{-3} \delta(\gamma(w)-y) e^{4\rho+i\sigma+2iH} \ .
\end{equation}
Now we note that since
\begin{equation}
	\partial_w \delta(\gamma(w)-x) = -[\partial \gamma(w)] \partial_x \delta(\gamma(w)-x) \ ,
\end{equation}
we have
\begin{equation} \label{eq:chain-rule-2}
	\partial_w^2 \delta(\gamma(w)-x) = -[\partial^2 \gamma(w)] \partial_x \delta(\gamma(w)-x) + [\partial\gamma(w)]^2 \partial_x^2 \delta(\gamma(w)-x) \ .
\end{equation}
Using this and eq.~\eqref{eq:delta-one-over} for $|x|>|y|$, we get
\begin{equation} \label{app:example-gpgm-1}
	P[\mathcal{G}^+,\mathcal{G}^-](x,y;w) \sim \frac{\mathcal{T}_{P=-4}(y,w)}{x-y} + \frac{\mathcal{A}^{\prime}(y,w)}{(x-y)^2}+\frac{2\mathcal{B}(y,w) [\partial\gamma(w)]^2}{(x-y)^3}\ ,
\end{equation}
where
\begin{equation} \label{eq:a-prime}
	\mathcal{A}^{\prime}(y,w) = \mathcal{A}(y,w) + [\partial^2 \gamma] \mathcal{B}(y,w) \ .
\end{equation}
Let us write the fields associated with the Cartan generator of the R-symmetry currents and the $w=1$-ground state in picture $P=-4$ following \cite{Gaberdiel:2021njm}
\begin{equation} \label{eq:j3-state}
	\mathcal{J}^3_{P=-4}(y,w)=\tfrac{1}{2} e^{2if_1-2if_2} (\partial \gamma)^{-2} \delta(\gamma(w)-y) e^{4\rho+i\sigma+2iH} \partial[if_1+if_2]\ ,
\end{equation}
\begin{equation} \label{eq:omega-p=-4}
	\Omega_{P=-4}(y,w) = e^{2if_1-2if_2} (\partial \gamma)^{-1} \delta(\gamma(w)-y) e^{4\rho+i\sigma+2iH}  \ ,
\end{equation}
see eq.~\eqref{eq:ground-state-pictures}. Having this, one can check that
\begin{align}
	\mathcal{A}^{\prime}(y,w) - 2\mathcal{J}^3_{P=-4}(y,w)= &-e^{2if_1-2if_2} (\partial \gamma)^{-2} \delta(\gamma(w)-y) e^{4\rho+i\sigma+2iH} \partial[if_1-if_2+2\rho+iH] \nonumber\\&+ [\partial^2 \gamma] e^{2if_1-2if_2} (\partial \gamma)^{-3} \delta(\gamma(z)-y) e^{4\rho+i\sigma+2iH} \ .
\end{align}
Let us define
\begin{equation}
	S(y,w) = -\tfrac{1}{2} e^{2if_1-2if_2} e^{4\rho+2iH}(\partial \gamma)^{-2} \delta(\gamma(w)-y) \ , \quad \tilde{G}^+_0 S(y,w)=0 \ ,
\end{equation}
where $S$ is primary and has weight $0$ on the world-sheet. By directly calculating the action of $G^+_0$, see eqs.~\eqref{eq:hybrid-n=2}, we can see that
\begin{equation} \label{eq:app-example-1}
	G^+_0 S(y,w) = \mathcal{A}^{\prime}(y,w)-2\mathcal{J}^3_{P=-4} + \tfrac{1}{2} e^{2if_1-2if_2} e^{4\rho+i\sigma+2iH}(\partial \gamma)^{-1} \partial_y \delta(\gamma(w)-y) \ ,
\end{equation}
where we have also used the chain rule in eq.~\eqref{eq:chain-rule}. The last term in eq.~\eqref{eq:app-example-1} is therefore equal to
\begin{equation}
	\tfrac{1}{2} J^+_0 \Omega_{P=-4} \ ,
\end{equation}
which is BRST exact, see \cite{Naderi:2022bus}. This shows that
\begin{equation} \label{eq:a-prime-j3}
	\mathcal{A}^{\prime}(y,w)-2\mathcal{J}^3_{P=-4}(y,w) = \text{BRST exact} \ .
\end{equation}
Moreover, the third pole in eq.~\eqref{app:example-gpgm-1} is equal to $2\Omega_{P=-4}(y,w)$. Therefore, the final answer up to BRST exact terms is
\begin{equation}
	P[\mathcal{G}^+,\mathcal{G}^-](x,y;w) \sim \frac{\mathcal{T}_{P=-4}(y,w)}{x-y} + \frac{2\mathcal{J}^{3}_{P=-4}(y,w)}{(x-y)^2}+\frac{2 \Omega_{P=-4}(y,w)}{(x-y)^3} \ ,
\end{equation}
which is the expected space-time answer, see Appendices~\ref{app:sca} and \ref{app:t4}.
%%%%%%%%%%%%%%%%%%%%%%%%%%
\subsection{Stress-tensor}
We begin by calculating the OPE of the topologically twisted stress-tensor in eq.~\eqref{eq:topologically-twisted-state} with itself
\begin{equation} \label{eq:app-tt}
	P[\mathcal{T}_{P=-4},\mathcal{T}_{P=-4}](x,y;w) \ .
\end{equation}
Recall that in eq.~\eqref{eq:d-stress-tensor}, $l_j$ is defined as
\begin{equation}
	l_j = if_2-\rho-iH^j \ , \quad j\in\{1,2\} \ ,
\end{equation}
which implies
\begin{equation}
	\partial l_j(z) \partial l_k(w) \sim \frac{\delta_{jk}}{(z-w)^2} \ .
\end{equation}
This shows that $D$ in eq.~\eqref{eq:d-stress-tensor} is the stress-tensor associated to the bosons $\partial X^j$ and $\partial \bar{X}^j$, and the bosons $l_j$ which have non-zero background charges. In fact, we have
\begin{equation} \label{eq:app-d-ope}
	D(z) D(w) \sim \frac{2 D(w)}{(z-w)^2} + \frac{\partial D(w)}{(z-w)} \ .
\end{equation}
We will use this to calculate the OPE in eq.~\eqref{eq:app-tt}. We begin by defining
\begin{equation} \label{eq:app-f}
	F(z)= e^{2if_1-2if_2} (\partial \gamma)^{-3} e^{4\rho+2iH} \ .
\end{equation}
Using eq.~\eqref{eq:app-d-ope} and a logic similar to the previous OPE calculations, see e.g.\ eqs.~\eqref{eq:delta-one-over}, \eqref{eq:delta-impose} and \eqref{eq:chain-rule}, we get
\begin{equation}
	P[\mathcal{T}_{P=-4},\mathcal{T}_{P=-4}](x,y;w) \sim \frac{\mathcal{U}(y,w)}{(x-y)}+\frac{\mathcal{V}(y,w)}{(x-y)^2} \ ,
\end{equation}
where we have defined
\begin{equation}
	\mathcal{U}(y,w)= e^{i\sigma} [2 F \partial F D+F^2 \partial D] \delta(\gamma(w)-y) \ ,
\end{equation}
\begin{equation}
	\mathcal{V}(y,w) = 2 e^{i\sigma} F^2 (\partial \gamma) \delta(\gamma(w)-y) D \ .
\end{equation}
Let us write the vertex operator in eq.~\eqref{eq:topologically-twisted-state} in picture $P=-8$
\begin{equation} \label{eq:app-t-p=-8}
	\mathcal{T}_{P=-8}(x,z) = e^{4if_1-4if_2} (\partial \gamma)^{-5} \delta(\gamma(z)-x) e^{8\rho+i\sigma+4iH} D \ . 
\end{equation}
In other words, we have
\begin{equation}
	P_+^2 \mathcal{T}_{P=-8}(x,z) = \mathcal{T}_{P=-4}(x,z) \ ,
\end{equation}
see Appendix~\ref{app:hybrid}. We note that we can write
\begin{equation} \label{eq:app-t-p=-8-rel}
	\mathcal{T}_{P=-8}(x,z)=e^{i\sigma} F^2(z) (\partial \gamma) \delta(\gamma(z)-x) D \ ,
\end{equation}
and therefore, the second pole is equal to $2\mathcal{T}_{P=-8}(y,w)$. In order to understand the first pole, we note that
\begin{equation} \label{eq:app-v1}
	\mathcal{U}(y,w)=e^{i\sigma}\partial[F^2 D \delta(\gamma(w)-y)]-e^{i\sigma}F^2 D \partial_w \delta(\gamma(w)-y) \ .
\end{equation}
The first term in eq.~\eqref{eq:app-v1} is BRST exact. To see this, let us set
\begin{equation} \label{eq:app-brst-exact}
	S(y,w) = F^2 D \delta(\gamma(w)-y) \ .
\end{equation}
$S$ has weight $0$ and is primary, and is also annihilated by $\tilde{G}^+_0$. Moreover, it is straightforward to check that
\begin{equation}
	G^+_0 S(y,w) = e^{i\sigma}\partial[F^2 D \delta(\gamma(w)-y)] \ .
\end{equation}
The second term in eq.~\eqref{eq:app-v1}, using eq.~\eqref{eq:chain-rule}, equals
\begin{equation}
	-e^{i\sigma}F^2 D \partial_w \delta(\gamma(w)-y) = e^{i\sigma} F^2 D (\partial \gamma) \partial_y \delta(\gamma(w)-y) = \partial_y \mathcal{T}_{P=-8}(y,w) \ ,
\end{equation}
where we have used eq.~\eqref{eq:app-t-p=-8-rel}. Therefore, up to BRST exact states, we have
\begin{equation}
	P[\mathcal{T}_{P=-4},\mathcal{T}_{P=-4}](x,y;w) \sim \frac{2 \mathcal{T}_{P=-8}(y,w)}{(x-y)^2}+ \frac{\partial_y \mathcal{T}_{P=-8}(y,w)}{(x-y)} \ ,
\end{equation}
which is the space-time answer that we were looking for.

In order to conclude the calculation of the OPE of the stress-tensor with itself, we have to also calculate the OPE of the Cartan generator of the R-symmetry currents, $\mathcal{J}^3$, with the topologically twisted stress-tensor, see eqs.~\eqref{eq:j3-state} and \eqref{eq:topologically-twisted-state}. We find it more convenient to work with $\mathcal{J}^3$ in picture $P=0$
\begin{equation} \label{eq:j3-p=0}
	\mathcal{J}^3_{P=0}(x,z) = K^3(z) e^{i\sigma} \delta(\gamma(z)-x) \ .
\end{equation}
Having this, an almost identical calculation to Section~\ref{sec:bosons} shows
\begin{equation} \label{eq:app-j3j3}
	P[\mathcal{J}^3_{P=0},\mathcal{J}^3_{P=0}](x,y;w) \sim \frac{\Omega_{P=0}(y,w)}{2(x-y)^2} \ ,
\end{equation}
which agrees with the space-time result, see Appendix~\ref{app:t4}.

Let us continue with the calculation of the OPE of $\mathcal{J}^3$ with $\mathcal{T}$. It is useful to calculate
\begin{equation}
	K^3(z) D(w) \sim \frac{1}{(z-w)^3}+\frac{\partial[2if_2-2\rho-iH]}{2(z-w)^2} \ ,
\end{equation}
see eq.~\eqref{eq:d-stress-tensor}. Using this and following the same logic in calculating the OPEs as in the previous cases (see e.g.\ eq.~\eqref{eq:chain-rule-2}), we get
\begin{align}
	P[\mathcal{J}^3_{P=0},\mathcal{T}_{P=-4}](x,y;w) &\sim \frac{e^{i\sigma}(\partial \gamma)^2 F \delta(\gamma(w)-y)}{(x-y)^3} \nonumber\\&+\frac{e^{i\sigma}\{(\partial^2 \gamma) F + F (\partial\gamma) \partial[2if_2-2\rho-iH]\}\delta(\gamma(w)-y)}{2(x-y)^2} \ ,
\end{align}
see eq.~\eqref{eq:app-f} for the definition of $F$. The third pole is equal to $\Omega_{P=-4}(y,w)$, see eq.~\eqref{eq:omega-p=-4}. For the double pole, we use that it equals $\frac{1}{2} \mathcal{A}^{\prime}$, see eq.~\eqref{eq:a-prime}. Now we use eq.~\eqref{eq:a-prime-j3} to conclude
\begin{equation} \label{eq:app-j-t}
	P[\mathcal{J}^3_{P=0},\mathcal{T}_{P=-4}](x,y;w) \sim \frac{\Omega_{P=-4}(y,w)}{(x-y)^3}+ \frac{\mathcal{J}^3_{P=-4}(y,w)}{(x-y)^2} \ ,
\end{equation}
up to BRST exact terms. Note that this is in agreement with the space-time result, see Appendix~\ref{app:t4}. Now we can use eq.~\eqref{eq:app-exchange} to conclude
\begin{equation} \label{eq:app-t-j}
	P[\mathcal{T}_{P=-4},\mathcal{J}^3_{P=0}](x,y;w) \sim \frac{-\Omega_{P=-4}(y,w)}{(x-y)^3}+ \frac{\mathcal{J}^3_{P=-4}(y,w)}{(x-y)^2} +\frac{\partial_y\mathcal{J}^3_{P=-4}(y,w)}{(x-y)}\ ,
\end{equation}
where we have expanded the delta-function $\delta(\gamma(w)-x)$ around $y$ by writing $x=y+(x-y)$, see eq.~\eqref{eq:taylor-formula}. This is also obviously in agreement with the space-time result.

The vertex operator associated with the stress tensor is
\begin{equation}
	\mathcal{L} = \mathcal{T} - \partial \mathcal{J}^3 \ ,
\end{equation}
see Appendix~\ref{app:sca}. Using the last property of eq.~\eqref{eq:p-product-arg} discussed in Appendix~\ref{app:ope}, we can calculate the OPE of $\partial \mathcal{J}^3$ with itself and also with $\mathcal{T}$ using the OPEs that we have calculate so far. Therefore, we conclude
\begin{equation}
	P[\mathcal{L}_{P=0},\mathcal{L}_{P=-4}](x,y;w) \sim \frac{3\Omega_{P=-4}(y,w)}{(x-y)^4}+\frac{2\mathcal{L}_{P=-4}(y,w)}{(x-y)^2}+\frac{\partial_y\mathcal{L}_{P=-4}(y,w)}{(x-y)} \ ,
\end{equation}
up to BRST exact states. This is the expected space-time answer, see Appendix~\ref{app:t4}.
%%%%%%%%%%%%%%%%%%%%%%%%%
\subsection{An explicit example of our argument} \label{app:explicit-arg}
We discuss with an example the condition in eqs.~\eqref{eq:locality-assumption} and \eqref{eq:s-f-modes}. Consider the physical vertex operators
\begin{align}
	\mathcal{A}(x,z) &= V([\partial \mathcal{X}^1_{-1}]^3 \ket{\Omega_{P=0}};x,z) \ , \\
	\mathcal{B}(y,w) &= V(\partial \bar{\mathcal{X}}^1_{-1} \ket{\Omega_{P=0}};y,w) \ .
\end{align}
We want to calculate the singular part of the corresponding OPE
\begin{equation} \label{eq:app3-to-calculate}
	P[\mathcal{A},\mathcal{B}](x,y;w) = \sum_{n \in \mathbb{Z}} V(\mathcal{A}_n \ket{\mathcal{B}};y,w) (x-y)^{-n-3} \ .
\end{equation}
First we calculate
\begin{equation}
	P[\partial \bar{\mathcal{X}}^1,\mathcal{A}](x,y;w) = \sum_{n \in \mathbb{Z}} V(\partial \bar{\mathcal{X}}^1_n [\partial \mathcal{X}^1_{-1}]^3 \ket{\Omega_{P=0}};y,w) (x-y)^{-n-1} \ .
\end{equation}
Using the algebra in eq.~\eqref{eq:ddf-algebra} and also using eq.~\eqref{eq:locality-assumption}, the singular part of this OPE is equal to
\begin{equation}
	P[\partial \mathcal{X}^1,\mathcal{A}](x,y;w) \sim \frac{3 V([\partial \mathcal{X}^1_{-1}]^2 \ket{\Omega_{P=0}};y,w)}{(x-y)^2} \ .
\end{equation}
Substituting this back into eq.~\eqref{eq:anti-comm-s-t}, we get
\begin{equation}
	[\partial \bar{\mathcal{X}}^1_n,\mathcal{A}_m] = 3 n \, \mathcal{C}_{n+m} \ , \quad \mathcal{C}(y,w) = V([\partial \mathcal{X}^1_{-1}]^2 \ket{\Omega_{P=0}};y,w) \ .
\end{equation}
Similarly, it is straightforward to see that $\mathcal{A}_n$ commutes with $\partial \mathcal{X}^1_m$. From eq.~\eqref{eq:app3-to-calculate}, we see that the singular part comes from $n\geq -2$. We get
\begin{equation}
	\mathcal{A}_n \partial \bar{\mathcal{X}}^1_{-1} \ket{\Omega_{P=0}} = 3 \mathcal{C}_{n-1} \ket{\Omega_{P=0}} + \partial \bar{\mathcal{X}}^1_{-1} \mathcal{A}_n  \ket{\Omega_{P=0}} \ .
\end{equation}
Using eq.~\eqref{eq:locality-assumption}, the second term is zero on cohomology if $n\geq -2$. The space-time weight of $\mathcal{C}$ is $h_C=2$, and so using eq.~\eqref{eq:locality-assumption}, the first term is non-zero only if $n \leq -1$. So we should only consider the cases where $n=-2$ or $n=-1$. If $n=-1$, we have
\begin{equation}
	\mathcal{C}_{-2} \ket{\Omega_{P=0}} = \ket{\mathcal{C}} = [\partial \mathcal{X}^1_{-1}]^2 \ket{\Omega_{P=0}} \ ,
\end{equation}
where we first used eq.~\eqref{eq:locality-assumption} and then the definition of $\mathcal{C}$. For $n=-2$, we use eq.~\eqref{eq:app-sl-derivative}
\begin{equation}
	\mathcal{C}_{-3} \ket{\Omega_{P=0}} = \mathcal{L}_{-1} \mathcal{C}_{-2} \ket{\Omega_{P=0}} \ .
\end{equation}
Therefore, using eq.~\eqref{eq:derivative-state}, for the singular part of the OPE in eq.~\eqref{eq:app3-to-calculate} we get
\begin{equation}
	P[\mathcal{A},\mathcal{B}](x,y;w) \sim \frac{3\mathcal{C}(y,w)}{(x-y)^2} + \frac{3 \partial_y \mathcal{C}(y,w)}{(x-y)} \ .
\end{equation}
Let us now compare this with the space-time OPE which is
\begin{equation}
	A(x) B(y) \sim \frac{3 \partial X^1(x) \partial X^1(x)}{(x-y)^2} \ .
\end{equation}
This shows that the OPE of $\mathcal{A}$ and $\mathcal{B}$ agrees with the space-time result.
%%%%%%%%%%%%%%%%%%%%%%%%%%%
\bibliographystyle{JHEP.bst}
\bibliography{bib.bib}
\end{document}